\documentclass[a4paper, 11pt]{article}
\usepackage{amsmath,amssymb,xcolor,comment}

\usepackage{jheppub} 

\usepackage[T1]{fontenc} 
\usepackage{framed}

\title{On the equivalence of unitarization prescriptions for the Sommerfeld enhancement}

\author{Barry E. Cimring,}
\author{Tracy R. Slatyer}

\affiliation{Center for Theoretical Physics -- a Leinweber Institute, Massachusetts Institute of Technology, Cambridge, MA 02139, USA}

\emailAdd{bcimring@mit.edu}
\emailAdd{tslatyer@mit.edu}

\abstract{The annihilation of self-interacting dark matter with long-range interactions can be significantly enhanced at low velocities through the Sommerfeld effect. At special points in parameter space, where near-zero-energy resonances exist in the spectrum of the theory, the standard calculation of this enhancement appears to violate unitarity. Recently, several approaches have been proposed to regulate this behavior, some introducing explicit ultraviolet (UV) scales and others not, raising the question of whether these prescriptions are consistent. In this paper, we compare these approaches and show that even in nominally cutoff-dependent methods, the regulated Sommerfeld-enhanced cross sections are independent of the UV regulator to a good approximation, and that when the unitarity-preserving corrections are large, the regulation schemes coincide to leading order. We use these insights to write down a regulator-independent prescription for unitarization applicable to multi-state systems, where the modified enhancement can be written solely in terms of the standard enhancement factor, the hard annihilation amplitude, and the $S$-matrix for scattering in the long-range potential.
}

\newcommand{\draftnote}[1]{} 

\begin{document} 

\begin{flushright}
MIT-CTP/6026
\end{flushright}
\maketitle
\flushbottom

\section{Introduction}

In the presence of a long-range attractive (repulsive) potential, short-range interactions between particles can be significantly enhanced (suppressed). This is termed the Sommerfeld effect, following early studies by Sommerfeld~\cite{Sommerfeld:1931qaf} of the distortion of the electron wave function by the Coulomb potential. Classically, this effect becomes significant when the potential energy of the interacting particles exceeds their kinetic energy; consequently, for weakly-coupled interactions, this is a low-velocity effect that can generally be treated in the framework of non-relativistic quantum mechanics. At leading order, the enhancement (or suppression) can be characterized by the behavior of the two-particle wavefunction as the interparticle separation $r$ goes to zero.

It was realized in 2003~\cite{Hisano:2003ec} that in the context of dark matter (DM) physics, there can be a broader range of phenomena associated with the Sommerfeld effect, and the long-range potential can be either mediated by Standard Model particles~\cite{Hisano:2003ec,Hisano:2004ds} or by entirely new forces (e.g.~some early references include Refs.~\cite{March-Russell:2008klu,ArkaniHamed:2008qn,Pospelov:2008jd}). The Sommerfeld enhancement can in principle be relevant for both the initial and final states, but the usual assumption in the context of DM annihilation is that it is relevant only for the non-relativistic initial-state DM particles, whereas the final-state particles are relativistic (see Ref.~\cite{Abe:2026blv} for a recent example where this assumption does not hold). We will adhere to this assumption in this work, and focus on initial-state effects.

In particular, when the two-body (initial-state) potential has a finite range (rather than the infinite-range Coulomb potential), at special points in parameter space there are additional resonant enhancements associated with the presence of zero- or finite-energy resonances in the spectrum (see e.g.~\cite{Kamada:2023iol,Beneke:2024iev} for discussion). In the standard method for computing the Sommerfeld enhancement, as introduced in Ref.~\cite{Hisano:2004ds}, the wavefunction is computed without accounting for the annihilation itself, using only the long-range potential. However, this prescription can lead to an enhanced partial-wave annihilation cross section which appears to violate unitarity at low velocities.

Specifically, partial-wave unitarity requires that for distinguishable particles, the total interaction cross section must obey:
\begin{align} \sigma_\ell(p) \le 4\pi(2\ell+1)/p^2, \label{eq:unitarity} \end{align}
where $p$ is the magnitude of the 3-momentum of either particle in the center-of-mass frame. The limit for inelastic processes (including annihilation) is lower by a factor of 4; for indistinguishable particles, the limit is increased by a factor of 2 (for those partial waves where selection rules allow a two-particle state). In the presence of resonant Sommerfeld enhancement, the low-velocity scaling is given by $\sigma_{\text{ann},\ell} \propto p^{-3}$ for $\ell=0$, and $\sigma_{\text{ann},\ell} \propto p^{2\ell-5}$ for $\ell \ge 1$ \cite{Kamada:2023iol}. Unitarity is thus generically violated on these resonances, at sufficiently low velocities, for partial waves $\ell \le 1$.

The origin of the unitarity violation is that the wavefunction of the interacting particles formally must include corrections due to the annihilation process, and when the annihilation cross section approaches the unitarity bound, these corrections are large \cite{Blum:2016nrz}. Ref.~\cite{Blum:2016nrz} proposed a method to incorporate these corrections, valid for the $\ell=0$ partial wave, by modeling the annihilation physics by a $\delta$-function potential with complex coefficient, and non-perturbatively solving for the modified wavefunction. This case has also been studied more recently in finite-temperature field theory \cite{Binder:2026fwe}.

More recently, three groups have proposed generalizations of the vacuum calculation to higher partial waves \cite{Flores:2024sfy,Parikh:2024mwa,Watanabe:2025kgw}. 
While all three methods ensure cross sections that manifestly obey the unitarity bound, the degree to which they are equivalent (or equivalent within some set of approximations) is not obvious. Refs.~\cite{Parikh:2024mwa,Watanabe:2025kgw} both require a hierarchy of scales between the range/energy scale of the annihilation process, and the range/energy scale of the long-range potential; the work of Ref.~\cite{Flores:2024sfy} (extended in Ref.~\cite{Flores:2025uoh}) does not impose this requirement and thus generalizes to situations where there is no large hierarchy, e.g. where the ``annihilation'' corresponds to capture into a bound state with radius comparable to the range of the potential \cite{Flores:2026yay}. Moreover, the result of Ref.~\cite{Parikh:2024mwa} nominally depends on a matching radius / ultraviolet (UV) cutoff parameter, while Ref.~\cite{Watanabe:2025kgw} requires independence of the cutoff as part of their calculation. Of these three methods, only that of Ref.~\cite{Parikh:2024mwa} covers the case where the potential couples multiple interacting 2-body states; this is relevant for calculations for heavy weakly-interacting DM and DM inhabiting non-Abelian dark sectors, where the DM is one state in a larger multiplet. 

Our goal in this work is to understand the relationship between these methods and to demonstrate their equivalence in the case where they share assumptions, i.e.~when there is only a single DM state, and there is a hierarchy between the short-range/high-energy physics associated with annihilation and the non-absorptive physics encoded in the long-range potential. We will develop a precise dictionary that demonstrates how to obtain the results of Ref.~\cite{Parikh:2024mwa} from those of Ref.~\cite{Flores:2025uoh} when these assumptions hold, which in particular clarifies that UV-divergent terms identified in the analysis of Ref.~\cite{Flores:2025uoh} are absorbed into the short-range amplitude in Ref.~\cite{Flores:2025uoh} (and correspondingly also Ref.~\cite{Watanabe:2025kgw}). We will demonstrate properties of the regular and irregular wavefunctions close to resonance that allow the result of Ref.~\cite{Parikh:2024mwa} to be approximated in a manifestly cutoff-independent form matching that of Ref.~\cite{Watanabe:2025kgw}, which involves only the regular solution to the long-range wavefunction, rather than the (divergent at the origin) irregular solution. Finally, we use the insights gained in the single-state matching to generalize the result of Ref.~\cite{Watanabe:2025kgw} to the multi-state case.

In section~\ref{sec:regulating_methods} we review the three unitarization methods we consider and introduce the notation that we will use throughout the paper. In section~\ref{sec:equivalence_singlestate} we demonstrate the (approximate) equivalence of these methods in the single-state case. In section~\ref{sec:multistate} we derive a simplified version of the multi-state result of Ref.~\cite{Parikh:2024mwa}, which generalizes the prescription of Ref.~\cite{Watanabe:2025kgw} to the multi-state case, and validate this prescription by numerical comparison between the methods for the case of wino DM. We present our conclusions in section~\ref{sec:conclusions}.

\section{Regulating methods for single-channel annihilation}
\label{sec:regulating_methods}
In this section, we outline the three existing approaches in the literature for regulating Sommerfeld-enhanced annihilation cross sections based on Refs.~\cite{Parikh:2024mwa,Flores:2025uoh,Watanabe:2025kgw}. This section will primarily be literature review, with the exception of Section~\ref{sec:irregular_regulator}, where we discuss the ``outgoing-wave regulator'' term that appears in all  three methods, which is named for reasons we will discuss later.  
We review these methods for only a single incoming two-particle state, so the potential is a scalar function (as opposed to a matrix), as this will be the case where we show the approaches discussed are equivalent under a certain set of assumptions. We similarly consider as our baseline the case of inclusive annihilation, or equivalently the case where only a single final state is relevant, as this facilitates the comparison. Refs.~\cite{Flores:2025uoh,Parikh:2024mwa} give expressions for exclusive regulated cross sections, and we will show how the matching we identify generalizes to this case, at least at leading order.

In the first of these approaches, as outlined in Ref.~\cite{Parikh:2024mwa}, the short-ranged annihilation physics is assumed to take place solely within a radius $r<a$, which we generally take to be comparable to the de Broglie wavelength of the annihilating particles $a \sim 1/\mu$, where $\mu$ is the reduced mass. Outside this radius, the particles are assumed to interact solely through long-ranged forces due to mediating particles, i.e. interactions that give rise to Hermitian potentials. We denote this potential as $V_L(r)$, and in principle it may be computed to any order; for practical examples in this work, we will generally use only the tree-level potential. Higher-order corrections to the potential can certainly change the energies of bound states and the positions of resonance peaks, and so may have a large impact on the cross section at any specific parameter point, but do not generally change the qualitative behavior of the theory (\cite{Beneke:2019qaa,Urban:2021cdu}; the effect of using the leading-order vs next-to-leading-order potential for unitarization is discussed in Ref.~\cite{Parikh:2024mwa}).  Note that in any case, we expect this non-relativistic potential to be valid only for $r \gtrsim 1/\mu$ (at shorter distances, in principle, the relativistic theory should be employed). We will refer to this position-space matching method as the PSS24 approach (Parikh, Sato, \& Slatyer~\cite{Parikh:2024mwa}), after the authors.

To compute the scattering amplitude under these assumptions, we match the asymptotic scattering solutions to the solutions in the short-distance region at the radius $r=a$. At the matching radius, if the wavefunction is decomposed into a ``plane-wave-like'' component and a ``purely outgoing'' component, the relative strength of the outgoing component is determined by the short-distance scattering amplitude. This amplitude includes annihilation, elastic scattering, and any other physics occurring at scales $r < a$. Thus the boundary conditions at $r=a$ encode the short-distance physics. We then solve the Schr\"{o}dinger equation with these modified boundary conditions.

Note that the short-range scattering amplitude can be decomposed into a part originating from the long-range potential $V_L(r)$, extended into the $r < a$ region, and an extra piece that is not accounted for in the long-range potential (which must include any absorptive physics). There is flexibility in this decomposition when it comes to short-range elastic scattering, which may be captured either in the short-distance behavior of $V_L(r)$ or in the additional non-potential contribution to the short-distance scattering amplitude. The final $S$-matrix does not depend on this decomposition, but varying this convention modifies both $V_L(r)$ and the non-potential contribution to the scattering amplitude, and thus can change intermediate steps in the calculation. 

In the second approach, given in Ref.~\cite{Watanabe:2025kgw}, it is understood that the short-range annihilation amplitude represents a secular term in the perturbative expansion of the scattering amplitude between the annihilating particles. 
Just as secular terms in renormalized perturbation theory can be re-summed with renormalization group- (RG-) improved perturbation theory, one can re-sum the perturbation series for the amplitude to obtain an RG-improved scattering amplitude. The RG-improved scattering amplitude must then be matched to a UV theory at an appropriate scale (which, for this problem, is $\sim\Lambda_{\text{QM}}$).  We will refer to the method of RG-improved perturbation theory as the W25 (Watanabe) approach.

The final approach of Ref.~\cite{Flores:2025uoh} (following earlier work \cite{Flores:2024sfy}) includes both the long-range Hermitian and short-ranged anti-Hermitian effects into the computation of the two-particle wavefunctions. Here, the short-range and long-range potentials in the Schr\"odinger equation are obtained from the low-energy limits of the amplitudes that appear in the Bethe-Salpeter equation. 
The full wavefunctions are obtained via the method of Green's functions, where the short-ranged potentials are treated as sources to the Schr\"odinger equation in the presence of only the long-range potential. 
The annihilation amplitude is then computed using the full wavefunctions, going beyond the Born approximation. We refer to this as the FP25 (Flores \& Petraki) approach, indicating the authors.

In all three cases, the resulting $S$-matrices are non-unitary when restricted to the states available through elastic scattering (and captured in the QM calculation), allowing for the full $S$-matrix including annihilation to be unitary. Each calculation results in an annihilation cross section that manifestly obeys the unitarity bound.

\subsection{The PSS24 approach: separating short- and long-distance physics}
To obtain the unitarized cross section in the PSS24 approach, one first solves the scattering problem in the presence of the long-range non-relativistic potential $V_L(r)$:
\begin{align}
    \bigg(-\frac{d^2}{dr^2} + \frac{\ell(\ell+1)}{r^2} + 2\mu V_L(r) -p^2\bigg)u_\ell(r) = 0,
\end{align}
for the regular $F_\ell(r)$ and irregular $G_\ell(r)$ families of real scattering solutions. We normalize these solutions such that 
\begin{align}
F_\ell(r)&\rightarrow C_\ell s_\ell(pr) \nonumber \\
G_\ell(r)&\rightarrow C_\ell^{-1} c_\ell(pr) \label{eq:shortrangebcs}
\end{align}
at the origin, where $C_\ell$ is the usual (unregulated) Sommerfeld enhancement factor (with the Sommerfeld enhancement to the cross-section being given by $C_\ell^2$; note $C_\ell$ is real). Here $s_\ell(x) = x j_\ell(x)$ and $c_\ell(x)=-xn_\ell(x)$ are the free-particle solutions to the Schr\"odinger equation written in terms of spherical Bessel functions. At asymptotic infinity, the solutions have the form 
\begin{align}
F_\ell(r)&\rightarrow \sin(pr -\pi\ell/2  + \delta_\ell) \nonumber \\
G_\ell(r)&\rightarrow \cos(pr -\pi\ell/2  + \delta_\ell),\label{eq:longrangebcs}
\end{align}
where $\delta_\ell$ is the standard phase shift for elastic scattering. In the above we have chosen the irregular solution such that $G_\ell(r) + iF_\ell(r)$ is purely outgoing at infinity, i.e.
\begin{align}
\label{eq:FG_large_r_BC}
    G_\ell(r) + i F_\ell(r) \rightarrow (-i)^\ell e^{i(pr+\delta_\ell)}
\end{align}
for large $r$ (note that before imposing this boundary condition, the irregular solution is defined modulo the addition of the regular solution).

In what follows, it will be useful to decompose the scattering solutions via the method of variable phase \cite{PhysRevC.84.064308,Beneke:2014gja,Asadi:2016ybp,Parikh:2024mwa}. In general, we can write
\begin{align}
\label{eq:variable_phase}
    u_\ell(r) =\alpha(r)f_\ell(r) - \beta(r)g_\ell(r),
\end{align}
where $f_\ell(r)=s_\ell(p r)/\sqrt{p}$ and $g_\ell(r)=(c_\ell(p r) + is_\ell(p r))/\sqrt{p}$ are the partial-wave decompositions of plane and purely-outgoing waves, respectively (up to a $1/\sqrt{p}$ normalization factor which we choose for convenience). We fix $\alpha$ and $\beta$ by requiring 
\begin{align}
    u_\ell'(r) =\alpha(r)f'_\ell(r) - \beta(r)g'_\ell(r).
\end{align}
This in turn implies that we can write
\begin{align}
    \alpha(r)&=W_r[g_\ell, u_\ell]\\
    \beta(r)&=W_r[f_\ell, u_\ell],
\end{align}
where $W_r[x,y]=x(r)y'(r)-x'(r)y(r)$ is the Wronskian between the functions $x$ and $y$ evaluated at $r$. $g_\ell$ and $f_\ell$ are normalized such that $W_r[g_\ell,f_\ell]=1$. We note that if $x,y$ are solutions to the same Schr\"odinger equation, the Wronskian is constant $\frac{d}{dr}W_r[x,y]=0$.

To obtain the $S$-matrix, one assumes there exists a scattering amplitude $f_{s,\ell}$ that describes the short-distance physics, such that the solution at the matching radius $r=a$
has an outgoing component weighted by $f_{s,\ell}$ (relative to the plane-wave-like component)
\begin{align}
\label{eq:soln_less_than_single_state}
    u_{
    \ell}(r) \simeq \gamma_\ell\big(s_\ell(pr) + f_{s,\ell}(c_\ell(pr) + is_\ell(pr)\big).
\end{align}
In this case, we use the $\simeq$ symbol to indicate that the two sides match up to values + first derivatives. This is {\it not} a perturbative relation; $f_{s,\ell}$ is intended to describe the short-distance scattering to all orders, although in practice it is computed approximately using a perturbative treatment.

Applying this boundary condition at $r=a$ to the solution of the long-distance Schr\"{o}dinger equation, and reading off the coefficient of the outgoing wave as $r\rightarrow \infty$, Ref.~\cite{Parikh:2024mwa} derived the $S$-matrix for the full scattering problem (including the short-distance physics):
\begin{align}
S_\ell=S_{0,\ell}\bigg(1 + \frac{2ipC_\ell^2\bar f_{s,\ell}}{1 - ipC_\ell^2 \bar f_{s,\ell}-\bar f_{s,\ell}\bar Z_\ell}\bigg) = S_{0,\ell}\bigg(1 + \frac{2ipC_\ell^2}{\bar f_{s,\ell}^{-1} - ipC_\ell^2 - \bar{Z}_\ell}\bigg).\label{eq:PSS-Smatrix}
\end{align}
Here, $S_{0,\ell}=e^{2i\delta_\ell}$ is the $S$-matrix in the absence of annihilation (i.e.~where all the relevant physics is encoded in the long-range potential $V_L(r)$), and $\bar Z_\ell$ is a regulating term obtained in the matching, given by
\begin{align} \label{eq:ZbarEll_variable_phase}\bar{Z}_\ell = p^{-1/2}(Q_\ell^aC_\ell)\alpha_{ G_\ell}(a).\end{align}
Here $Q_\ell^a=C_\ell^ae^{i\delta_\ell^a}$, where $C_\ell^a$ is the (unregulated) Sommerfeld enhancement factor when $V_L(r)$ is set to zero outside the matching radius, and $\delta_\ell^a$ is the corresponding phase shift. Similarly, we define $Q_\ell = C_\ell e^{i\delta_\ell}\equiv  C^{a\rightarrow \infty}_\ell e^{i\delta^{a\rightarrow \infty}_\ell}$. 
$\alpha_{G_\ell}(a)$ here denotes the $\alpha(r)$ coefficient for the irregular $G_\ell(r)$ solution, evaluated at $r=a$, i.e.~$\alpha_{G_\ell}(a) = W_a[g_\ell, G_\ell]$. $\bar f_{s,\ell}$ is the scattering amplitude with the effect of $V_L(r)$ ``factored out'', related to the short-distance scattering amplitude in~(\ref{eq:soln_less_than_single_state}) by 
\begin{align}
\label{eq:bare_ann_amp}
    \bar f_{s,\ell}=(Q_\ell^a)^{-2}(f_{s,\ell}-f_{b,\ell}),
\end{align}
where $f_{b,\ell}$ is the amplitude for scattering  off the potential $V_L(r)$ which is set to zero outside the matching radius. The effect of subtracting $f_{b,\ell}$ is to remove the pure elastic-scattering contribution that would be present even in the absence of any additional short-range physics beyond $V_L(r)$, and the division by $(Q^a_\ell)^2$ serves to remove the Sommerfeld-enhancement-like prefactor due to evolution of the wavefunction within $r=a$. The imaginary part of $\bar{f}_{s,\ell}$ can be related to the pure annihilation cross section obtained from the UV theory (see Appendix D of Ref.~\cite{Parikh:2024mwa} for details on this matching), at least at leading perturbative order. 

Note that while $\bar{f}_{s,\ell}$ goes to zero exactly when the only short-range physics is well-described by $V_L(r)$ (due to the subtraction of $f_{b,\ell}$), and in this sense contains no ``pure elastic scattering'' contribution, it will generically include higher-order corrections that involve both ``annihilation'' and ``elastic scattering'' interactions in a mixed ladder diagram, and contributions that cannot be written as a ladder-type diagram. Systematically improving the matching for $\bar{f}_{s,\ell}$ would require evaluating these higher-order diagrams and keeping careful track of the effect of turning off interactions outside $r=a$ (which we would expect to lead to an $a$-dependence in the result); this would likely require a careful effective-field-theory treatment that goes beyond the scope of Ref.~\cite{Parikh:2024mwa}. We also generically expect the computation of $\bar{f}_{s,\ell}$ in the UV theory to require renormalization; Ref.~\cite{Parikh:2024mwa} assumes this renormalization has already been performed and $\bar{f}_{s,\ell}$ can be written as a finite function of renormalized parameters, so any higher-order matching for $\bar{f}_{s,\ell}$ would also require the computation of diagrams including counterterms. We will show later in this work that the matching between Ref.~\cite{Parikh:2024mwa}  and Ref.~\cite{Flores:2025uoh} involves terms which are nominally UV-divergent, consistent with this picture.

In what follows, we work in the convention that the long-range potential is independent of the matching radius (in particular, we do not set $V_L(r)$ to zero within $r=a$), and so $C_\ell$ is $a$-independent.
As a result, the regulating term $\bar Z_\ell$ is the only quantity that explicitly depends on the matching radius $a$. As discussed above, we expect higher-order terms in $\bar{f}_{s,\ell}$ to also carry $a$-dependence; the two sources of $a$-dependence should cancel in the final result for the scattering cross section. However, when the absorptive physics encoded in $\bar{f}_{s,\ell}$ is short-range, we expect the $a$-dependence in that term to be small; we will demonstrate for the first time in this work that the $a$-dependence in $\bar{Z}_\ell$ is likewise parametrically small.

$\bar Z_\ell$ also happens to be the only quantity that depends on the purely-outgoing irregular solution, and cannot be derived (in this approach) directly from the regular solution, and so we refer to $\bar Z_\ell$ as the {\it outgoing-wave regulator}. The annihilation cross section is obtained via:
\begin{equation}
\label{eq:ann_cross_section}
    (\sigma v_{\text{rel}})^{\text{ann.}}_\ell=c_i\frac{\pi}{\mu p}(2\ell+1)(1-|S_\ell|^2)
\end{equation}
and we have
\begin{equation}
\label{eq:PSS24_cross_section_ss}
(\sigma v_{\text{rel}})^{\text{ann.}}_\ell=c_i\frac{4\pi}{\mu }(2\ell+1) \frac{C_\ell^2\text{Im}(\bar f_{s,\ell})}{\Big|1-(ipC_\ell^2 + \bar Z_\ell)\bar f_{s,\ell}\Big|^2},
\end{equation}
where $c_i = 1\:(2)$ for distinguishable (identical) particles, and we have dropped higher-order terms in $\bar{f}_{s,\ell}$ in the numerator.

In the following sections, we will show that near a resonance, $\bar Z_\ell\bar f_{s,\ell}$ is dominated by a large, $a$-independent constant with negligible $a$-dependent corrections. Further, we will see that far from resonance, $\bar Z_\ell \bar f_{s,\ell}$ is small and negligible. The cross section will therefore be largely independent of the matching radius, which is expected on physical grounds,\footnote{To reiterate, on physical grounds, we expect the cross section to be {\it fully} independent of the matching radius once subleading $a$-dependent corrections to $\bar{f}_{s,\ell}$ are also taken into account, but this would require a higher-order matching to the relativistic theory, as well as keeping track of terms that we have neglected by assuming the physics {\it outside} $r=a$ is fully characterized by the non-relativistic potential $V_L(r)$.} necessary for consistency with the other treatments, and consistent with the numerical results of the PSS24 approach given in Ref.~\cite{Parikh:2024mwa}.

\subsection{The W25 approach: RG-improved annihilation amplitudes}
In the unitarization prescription of Ref.~\cite{Watanabe:2025kgw}, RG-improved perturbation theory is employed to re-sum the perturbative expansion of the scattering amplitude in the presence of the short-ranged absorptive and long-distance non-absorptive physics. The author notes that the existence of shallow bound states in the spectrum leads to secular terms in the scattering amplitude after integrating out heavy degrees of freedom. The secular terms result in a singular perturbation theory for the scattering amplitude. Following Ref.~\cite{Watanabe:2025kgw}, the perturbative expansion can be regularized using RG-improved perturbation theory, extending the validity of the amplitude to the low-velocity regime. 

To compute the RG-improved amplitude, one first computes the perturbative result for the amplitude in the distorted-wave Born approximation (DWBA). In the DWBA, the total scattering amplitude in the presence of the long-range potential and short-range perturbative physics is 
\begin{align}
\label{eq:W25_amplitude}
    f_\ell\simeq f_\ell^L +Q_\ell^2 f_\ell^S,
\end{align}
where $f_{\ell}^L=(e^{2i\delta_\ell}-1)/2ip$ is the scattering amplitude due to the long-range potential, and $f_\ell^S$ is the scattering amplitude due to the short-ranged anti-Hermitian potential $V_S(r)$
\begin{align}
\label{eq:W25_fS}
f_\ell^S =- \frac{1}{C_\ell^2}\frac{2\mu}{p^2} \int_{0}^{\infty} dr F_\ell(r) V_S(r)F_\ell(r),
\end{align}
which we assume has support within $r<a$ (note that $1/Q_\ell$ corresponds to the Jost function discussed in Ref.~\cite{Watanabe:2025kgw}). 

The first step in the RG-improved perturbation procedure is to introduce the appropriate power of the quantity $C(Q)Z(Q)=1$ into each term in the amplitude~(\ref{eq:W25_amplitude}), which is formally unity. Here, $Q$ is an arbitrary scale and $Z(Q)$ is a counter-term as in standard RG computations. One then expands $Z(Q)=1+Z_1(Q)$ and the amplitude to first-order, and demands $Z_1(Q)$ cancels completely the secular term due to the short-ranged physics at a given scale $p$. A renormalization group-like equation can then be obtained for $C(Q)$ by demanding the amplitude be $Q$-independent, i.e. $Q\frac{d}{dQ}f_\ell=0$. The result of solving this equation for $C(Q)$ is the RG-improved scattering amplitude:
\begin{align}
    f_\ell^{\text{imp}}\simeq \frac{f_\ell^L}{1-Q_\ell^2\frac{f_{\ell}^S}{f_\ell^L}},
\end{align}
which has been matched onto the Born approximation at the scale $p_0\sim\Lambda_{\text{QM}}$. The corresponding $S$-matrix is
\begin{align}
    S_\ell=1+2 ip f_{\ell}^\text{imp}= S_{0,\ell}\left(1+\frac{2ipC_\ell^2f_\ell^S}{1-ipC_\ell^2f_\ell^S-pC_\ell^2\cot\delta_\ell f_\ell^S}\right), \label{eq:W25_S_matrix}
\end{align}
where, again, $S_{0,\ell}=1 + 2ipf_\ell^L=e^{2i\delta_\ell}$ is the $S$-matrix due to the long-range potential. 
Finally, the cross sections can be obtained via the formula~(\ref{eq:ann_cross_section}), which we write in the suggestive form:
\begin{align}
\label{eq:W25_cross_section}
(\sigma v_{\text{rel}})^{\text{ann.}}_\ell=c_i\frac{4\pi}{\mu }(2\ell+1) \frac{C_\ell^2\text{Im}( f_{\ell}^S)}{\Big|1-(ipC_\ell^2 + pC_\ell^2\cot\delta_\ell) f_{\ell}^S\Big|^2}, 
\end{align}
after dropping higher-order terms in $f_{\ell}^S$ in the numerator. We note that the $pC_\ell^2\cot\delta_\ell$ term in the denominator of~(\ref{eq:W25_S_matrix}) (and consequently in (\ref{eq:W25_cross_section})) plays exactly the same ``outgoing-wave regulator'' role as $\bar Z_\ell$ in the PSS24 result (\ref{eq:PSS-Smatrix}), despite (unlike $\bar{Z}_\ell$) not depending on the outgoing irregular solution to the Schr\"{o}dinger equation. In particular,
if we could identify  $\bar{f}_{s,\ell} = f_\ell^S$, $p C_\ell^2 \cot \delta_\ell = \bar{Z}_\ell$, then the two results would exactly agree (at the level of the $S$-matrix and hence for all cross sections derived from the $S$-matrix).  We will show in this work that---while it is not true that $\bar{Z}_\ell = p C_\ell^2 \cot\delta_\ell$ for arbitrary parameters---there is a close relation between these two terms whenever this correction term is significant.

\subsection{The FP25 approach: computing the unitarized $S$-matrix via the self-energy kernel}
The computation of the 2PI self-energy kernel allows for a more diagrammatic calculation of the unitarized Sommerfeld-enhanced annihilation cross section, which can incorporate both Hermitian and anti-Hermitian physics into the two-body scattering problem. This approach was initially introduced in Ref.~\cite{Flores:2024sfy}, and then extended to elastic scattering and bound-state formation in Refs.~\cite{Flores:2025uoh,Flores:2026yay}. We will focus primarily on matching the results of Ref.~\cite{Flores:2025uoh} (which we denote FP25), which are more general than the results of Ref.~\cite{Flores:2024sfy}.

Following Ref.~\cite{Flores:2025uoh}, the Schr\"odinger equation is obtained in the limit of the instantaneous approximation applied to the resummation of the 2PI self-energy kernel. 
The annihilation amplitude is then obtained using the solutions to the Schr\"odinger equation in the presence of the 2PI self-energy kernel (which appears as a potential) as weights for the scattering plane waves. 

Assuming the long-range interactions are local and Hermitian, and, in the instantaneous approximation, that the absorptive interactions can be captured by a (non-central) optical potential that factors as $\mathcal{V}_\ell(r,r') \propto v_\ell^*(r') v_\ell(r)$  (as can be seen from the optical theorem), the Schr\"odinger equation obtained from the 2PI kernel takes the form:
\begin{align}
\label{eq:self_energy_SE}
    \bigg(-\frac{d^2}{dr^2}+\frac{\ell(\ell+1)}{r^2}+2\mu V_L(r)-p^2\bigg)u_\ell(r) =\eta_\ell^2 r v^*_\ell(r) \int_0^\infty dr' r' v_\ell(r')u_\ell(r'),
\end{align}
where $v_\ell(r)$ describes the potential profile associated with the annihilation physics, and $\eta_\ell^2$ is a complex coupling defined in Ref.~\cite{Flores:2025uoh}. In what follows, we assume there is a single channel for annihilation such that the short-ranged potential has only the term on the right-hand side above. We have changed the notation between that of Ref.~\cite{Flores:2025uoh} and the results presented here; Table~\ref{tab:FP_dict} provides the dictionary for the equivalent quantities in our paper.

\begin{table}[t]
\centering
\begin{tabular}{lcc}
\hline\hline
 & \textbf{Flores-Petraki}~\cite{Flores:2025uoh} & \textbf{This paper} \\
\hline
Wave function 
& $u_{p,\ell}(r)$ 
& $c_i^{1/2} i^\ell e^{i\delta_\ell}u_\ell(r)$ \\[4pt]

Long-range potential 
& $V(r)$ 
& $V_{L}(r)$ \\[4pt]

Short-range potential 
& $\nu_\ell(r)$ 
& $v_{\ell}(r)/\sqrt{2\mu}$ \\[4pt]

Phase-shift by long-range potential 
& $\theta_\ell(p)$ 
& $\delta_\ell$ \\[4pt]

Regular wave function 
& $\mathcal{F}_{p,\ell}(r)$ 
& $c_i^{1/2}i^\ell e^{i\delta_\ell} F_\ell(r)$ \\[4pt]

Irregular wave function 
& $\mathcal{G}_{p,\ell}(r)$ 
& $-\,c_i^{1/2}i^\ell e^{i\delta_\ell} G_\ell(r)$ \\[4pt]

Green's function 
& $G_{p,\ell}(r,r')$ 
& $ 2\mu \mathcal G_\ell(r,r')$ \\[4pt]

Unregulated amplitude 
& $ \hat{M}_{\ell,\rm unreg}(p)$ 
& $ i^\ell e^{i\delta_\ell}M^{\rm unreg.}_\ell(p)$ \\[4pt]

Regulated amplitude 
& $\hat{M}_{\ell,\rm reg}(p)$ 
& $ i^\ell e^{i\delta_\ell}M^{\rm reg.}_\ell(p)$ \\[4pt]

Regularization matrix 
& $\mathbb N_\ell(p)$ 
& $ N_\ell(p)$ \\[4pt]

Outgoing-wave regulator 
& $\mathbb W_\ell(p)$ 
& $-W_\ell(p)$ \\[4pt]

\hline\hline
\end{tabular}

\caption{Notation in Flores-Petraki~\cite{Flores:2025uoh}. The rightmost column shows the corresponding expression in the notation of this paper.}
 \label{tab:FP_dict}
\end{table}

Treating the right-hand side of~(\ref{eq:self_energy_SE}) as a source, the regular solution to the full Schr\"odinger equation in terms of regular and irregular solutions to the purely long-range Schr\"odinger equation is 
\begin{align}
    u_\ell(r) = F_\ell(r) + \eta_\ell^2 \int_0^\infty dr'   r' \mathcal{G}_\ell(r,r')v^*_\ell(r')\int_0^\infty dr'' r'' v_\ell(r'')u_\ell(r''),
\end{align}
where $\mathcal{G}_\ell(r,r')$ is the Green's function for the Schr\"odinger equation in the presence of only the long-range potential
\begin{align}
    \bigg(-\frac{d^2}{dr^2}+\frac{\ell(\ell+1)}{r^2}+2\mu V_L(r)-p^2\bigg)\mathcal G_\ell(r,r') = \delta(r-r').
\end{align}
Note the above differs from the full wavefunctions in Ref.~\cite{Flores:2025uoh} by a phase. The Green's function can be expressed in terms of the regular and irregular solutions according to 
\begin{align}
    \mathcal G_\ell(r,r') = \frac{ i}{p} F_\ell(r_<)(F_\ell(r_>) - i G_\ell(r_>)),
\end{align}
where $r_{<,>}=\min,\max (r,r')$. 
Inverting the above, the solution to the Schr\"odinger equation is 
\begin{align}
    u_\ell(r) = F_\ell(r) + \eta_\ell^2 \int_0^\infty dr'  r' \mathcal{G}_\ell(r,r')v^*_\ell(r')N_\ell(p)^{-1}\int_0^\infty dr'' r'' v_\ell(r'') F_\ell(r'')
\end{align}
where 
\begin{equation}
    N_\ell(p)=1 - \eta_\ell^2 \int r dr r' dr' v_\ell(r)\mathcal G_\ell(r,r')v^*_\ell(r')
\end{equation}
is referred to as the regulating matrix (which is $1\times 1$ in the single-annihilation channel case considered here).
The regulated amplitude is computed via 
\begin{align}
\label{eq:FP_regulated_amplitude}
    M^{\rm reg.}_\ell=\frac{1}{\sqrt{p}} \int_{0}^{\infty} dr r u_\ell(r) v_\ell(r).
\end{align}
Note this is a non-perturbative result and is {\it not} the first-order DWBA, since $u_\ell$ itself includes (non-perturbative) corrections from the short-range annihilation. 

The $S$-matrix is
\begin{align}
    S_\ell& = S_{0,\ell} \left(1 + \frac{2 i |M^{\rm unreg.}_\ell|^2 \eta_\ell^2/(1 - \eta_\ell^2 W_\ell)}{1 - i |M^{\rm unreg.}_\ell|^2 \eta_\ell^2/(1 - \eta_\ell^2 W_\ell)} \right)= S_{0,\ell}\left(1 + \frac{2 i \eta_\ell^2 \left|M^{\rm unreg.}_\ell\right|^2 }{1 - \eta_\ell^2 W_\ell - i \eta_\ell^2 \left|M^{\rm unreg.}_\ell\right|^2}\right) \label{eq:FP-Smatrix}
\end{align}
where $M^{\rm unreg.}_\ell$ is the unregulated annihilation amplitude computed via the DWBA
\begin{align}
\label{eq:FP_unregulated_amplitude}
    M^{\rm unreg.}_\ell=\frac{1}{\sqrt{p}} \int_{0}^{\infty} dr r F_\ell(r) v_\ell(r),
\end{align}
 and $W_\ell$ is the irregular part of the Green's function integral weighted by the annihilation potential
\begin{align}
\label{eq:FP_Wl_definition}
W_\ell=\frac{1}{p}\int_0^\infty dr r\int_0^\infty dr'r'F_\ell(r_<)G_\ell(r_>)v_\ell(r)v^*_\ell(r').
\end{align}
Note that $W_\ell$ may be divergent and so may require renormalization. For example, for a $\delta$-function absorptive potential with $\ell=0$ (studied in Ref.~\cite{Blum:2016nrz} and checked against the FP25 formalism in Ref.~\cite{Flores:2025uoh}), and the simplest case with no long-range potential at all, we have $v_0(r) \propto \delta(r)/r^2$, $F_0(r) \propto r$, $G_0(r)\propto r^0$ as $r\rightarrow 0$, and we see the integral has a divergent contribution from $r\rightarrow 0$. 

Finally, it is illustrative to consider the simplified case where the additional physics (beyond the long-range potential) is dominated by absorptive processes. In this case, we can write $\eta_\ell^2 \rightarrow i$ (absorbing any real prefactor into $v_\ell(r)$), in which case FP25 implies
\begin{align}
    (\sigma v_{\text{rel}})^{\text{ann.}}_\ell=c_i\frac{4\pi}{\mu}(2\ell+1)\frac{\left|M_\ell^{\rm unreg.}
    \right|^2}{\Big|1+\left|M_\ell^{\rm unreg.}
    \right|^2 - iW_\ell\Big|^2}.
\end{align}
Again, this result has an identical structure to (\ref{eq:ann_cross_section}, \ref{eq:W25_cross_section}), noting that in this case $|M_\ell^{\rm unreg.}|^2$ controls the unregulated Sommerfeld-enhanced annihilation cross section and so is expected to be proportional (at least to leading order) to $C_\ell^2 \text{Im}(\bar{f}_{s,\ell})$ appearing in (\ref{eq:ann_cross_section}). We note that the expression for the cross section here differs from the result of Ref.~\cite{Flores:2024sfy} by the inclusion of $W_\ell$, which involves the irregular solution to the Schr\"{o}dinger equation and plays a similar role in the cross-section formula to the outgoing-wave regulator of PSS24 (as given in (\ref{eq:ann_cross_section})). In Ref.~\cite{Flores:2024sfy}, simplifying assumptions were made to  express the Green's function in a form which did not contain the irregular solution, with the result that  their expression for the unitarized cross section does not include the term controlled by $W_\ell$. Here we do not make these assumptions, matching the final result of Ref.~\cite{Flores:2025uoh}, where these assumptions were also not made. 

\subsection{The outgoing-wave regulator}
\label{sec:irregular_regulator}

We have seen in the previous sections that all three methods give rise to very similar formulae for the $S$-matrix and corrected cross section. In the leading-order corrected cross section, the unregulated cross section is multiplied by a factor of the form $1/(1 + X)$ that prevents violation of unitarization. Here $X$ always includes a term proportional to the unregulated cross section (including the Sommerfeld enhancement), and it can be readily checked that this term  on its own is sufficient to ensure unitarization. However, each approach also includes in $X$ an ``outgoing-wave regulator'' (controlled by $\bar{Z}_\ell$ for PSS24, $\cot\delta_\ell$ for W25, and $W_\ell$ for FP25). The question of equivalence between these outgoing-wave regulators is thus the most non-trivial aspect of the equivalence between the three approaches (although the matching between the unregulated annihilation amplitudes, beyond leading order, also requires some care with regard to the FP25 case, and will be discussed in more detail in Section~\ref{sec:equivalence_singlestate}). In this subsection, we briefly discuss the phenomenological consequences and interpretation of this outgoing-wave regulator term, focusing on the PSS24 approach. 

In the absence of the $\bar{Z}_\ell$ outgoing-wave regulator, if $\bar{f}_{s,\ell}$ was purely imaginary (at least at leading order), then the regulated cross section given in (\ref{eq:PSS24_cross_section_ss}) would be solely determined by the unregulated cross section (via the combination $C_\ell^2 \text{Im}(\bar{f}_{s,\ell})$) and the momentum $p$. This is also the behavior found in Ref.~\cite{Flores:2024sfy} if certain convergence criteria are satisfied for the high-momentum behavior of the amplitudes. In particular, this would imply that when the unregulated cross section is well below unitarity, the regulated and unregulated cross sections will always agree.

\begin{figure}[t]
    \centering
\includegraphics[width=0.6\textwidth]{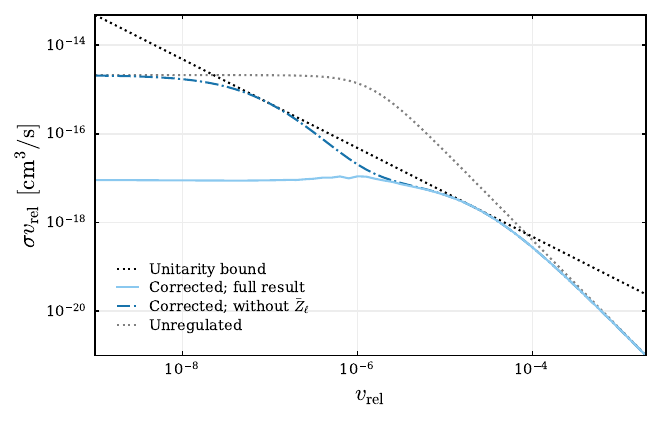}
    \caption{Example of the effect of the outgoing-wave regulator, using the formalism developed in Ref.~\cite{Parikh:2024mwa}. The dotted line indicates the velocity-dependent unitarity bound. The gray dotted line gives the unregulated, Sommerfeld-enhanced cross section; the light blue solid line gives the full regulated cross section; the dark blue dot-dashed line indicates the cross section one would obtain by setting the ``outgoing-wave regulator'' term controlled by $\bar{Z}_\ell$ to zero. The ``unregulated'' and ``corrected; full result'' lines in this plot are reproduced from Ref.~\cite{Parikh:2024mwa}; details of the input parameters may be obtained there.
    }
    \label{fig:v_dep_example}
\end{figure}

The presence of the outgoing-wave regulator can lead to large deviations from this behavior. An example of this behavior was already given in Ref.~\cite{Parikh:2024mwa}, for the case of $s$-wave wino DM annihilation, and is reproduced in Figure~\ref{fig:v_dep_example} (see Ref.~\cite{Parikh:2024mwa} for details of the inputs to this plot if desired; in the context of this work, we use it only as a qualitative example to help build intuition). In this case, close to the resonance but not exactly on it, the unregulated cross section (gray dotted line) can exceed the unitarity bound (black dotted line) at intermediate velocities, but then fall back below the unitarity bound at low velocities (exactly on a zero-energy resonance, in contrast, we generally expect the unregulated cross section to continue to grow relative to the unitarity bound as the velocity decreases, in cases where the resonant scaling led to unitarity violation in the first place). In the absence of the outgoing-wave regulator, we would expect the regulated cross section to converge to the unregulated cross section in this low-velocity regime, and this is indeed what occurs when we artificially set the outgoing-wave regulator to zero (dark blue dot-dashed line). However, in the full corrected solution (light blue solid line), instead the regulated cross section saturates at a value that can be orders of magnitude below the unregulated cross section (which is itself orders of magnitude below the unitarity bound).  In proximity to a resonance, the outgoing-wave regulator thus drives large differences in the predicted low-velocity behavior.

In the PSS24 formalism, the outgoing-wave regulator $\bar{Z}_\ell$ (given in \ref{eq:ZbarEll_variable_phase}) can be written (as shown in Ref.~\cite{Parikh:2024mwa}) in terms of Sommerfeld factors (both from the full potential, and the potential restricted to $r < a$), combined with the coefficient of $s_\ell(p r)$ in the variable phase expansion at $r=a$ of the complex solution $G_\ell(r) + i F_\ell(r)$. In other words, the information in $\bar{Z}_\ell$ that goes beyond Sommerfeld factors is encoded in the {\it ingoing} component, at $r=a$, of a wave that is purely outgoing at infinity (since the variable phase decomposition splits the wavefunction into a $s_\ell(p r)$-like part and a purely outgoing part). This contribution must consequently go to zero in the absence of scattering at $r > a$, and so we expect it to be controlled primarily by long-distance/infrared (IR) scattering in the potential, rather than the short-distance physics associated with the matching radius or the short-range divergence of the irregular solution. This intuition is consistent with the behavior of the W25 analogue of the outgoing-wave regulator, which is manifestly $a$-independent and can be determined directly from the regular solution.\footnote{We thank Yuki Watanabe for comments on this issue that helped clarify our intuition.} This picture also suggests that we can think of the outgoing-wave regulator $\bar{Z}_\ell$ as encoding the part of the outgoing wave, sourced by the short-distance physics, that scatters within the potential and consequently is reflected inward to smaller $r$ -- this may provide some additional intuition for $\bar{Z}_\ell$ becoming large close to resonance, although we do not rely on this idea in what follows. 

Strictly speaking, this picture of re-scattering is not unique to the $\bar Z_\ell$ term: the regulating term $pC_\ell^2\bar f_{s,\ell}$ in \eqref{eq:PSS24_cross_section_ss} can also be viewed as arising from the re-scattering of the outgoing wave, but this term is completely dictated by the bare Sommerfeld enhancement, and so takes no further IR information as input.

In the following section, we will make this intuition precise and prove the (approximate) equivalence between the PSS24 and W25 approaches. We will also show that the outgoing-wave regulator in the FP25 case can be split into two terms, one of which is solely controlled by the UV physics ($r < a$) and may be divergent, and the other of which is UV-finite but potentially enhanced by the long-range potential. The first term contributes to the higher-order matching of the short-range amplitude, while the second can be identified with the PSS24 (and W25) outgoing-wave regulator.  

\section{The equivalence of regulating approaches}
\label{sec:equivalence_singlestate}
Our aim now is to show that the methods of Section~\ref{sec:regulating_methods} for regulating Sommerfeld-enhanced annihilation yield approximately equivalent  $S$-matrices, with the corollary that the unitarized cross sections (for both elastic and inelastic processes) are approximately independent of the regulating prescription.
We find that near a zero- or finite-energy resonance, the regulators of the three approaches coincide up to terms parametrically suppressed in powers of the relative velocity, and are negligible far from resonance where they may or may not coincide. 

Before demonstrating consistency between the approaches considered, we first recall some of their obvious differences. Firstly, in the PSS24 and FP25 approaches, but not in the W25 approach, knowledge of the irregular solution is seemingly required to obtain the unitarized cross sections. Unlike the regular solution, the irregular solution is defined modulo the addition of the regular solution (before fixing its boundary conditions at infinity). 
We will see that near a resonance, the information in the irregular solution relevant for regulating the cross sections is precisely its boundary conditions at infinity, which is equivalent to information in the long-range unitary $S$-matrix.
Secondly, the PSS24 and W25 approaches explicitly separate the short- and long-distance physics (assuming some finiteness of the spatial extent of the perturbative physics), while the FP25 approach does not. 
Lastly, the result of the PSS24 approach contains explicit cut-off dependence, while the other approaches do not. 
Our aim will therefore be to relate the quantities in the PSS24 approach that appear to depend on the short-distance UV cutoff to the long-range physics, and to relate the result of the FP25 approach to the PSS24 approach when a separation of scales is allowed. 

Central to our discussion will be the effective-range expansion of the scattering phase-shift:
\begin{equation}
\label{eq:ERE_ss}
p^{2\ell+1}\cot\delta_\ell = \epsilon^{2\ell+1} + \frac{1}{2r_\ell^{2\ell-1}}p^2 + \mathcal{O}(p^4),
\end{equation}
which will allow us to quantify the proximity of the scattering system to a resonance. Here, $\epsilon$ is the inverse scattering length which vanishes on a zero-energy resonance. Although difficult to solve for in general, $\epsilon$ is nonetheless a useful notion for quantifying a ``distance'' from resonance in the parameter space of the scattering system~\cite{Kamada:2023iol, Kinugawa:2024jwq}. 

It will be useful in the following to note the difference in resonant behavior of the $p$-wave and higher partial cross sections as compared to the $s$-wave. As pointed out in Ref.~\cite{Beneke:2024iev}, finite-energy resonances can occur in $p$-wave or higher cross sections in the presence of an attractive potential, due to the potential barrier provided by the centrifugal term. In particular, when $\epsilon <0$ and $r_\ell >0$, there exist finite-energy resonances when $p^2\simeq-\epsilon^{2\ell+1}/r_\ell^{2\ell-1}$, corresponding to quasi-bound states in the spectrum. Quasi-bound states therefore do not exist for positive $\epsilon$. Quasi-bound states are also not present in $s$-wave scattering due to the lack of a centrifugal term. The relevant $s$-wave resonances are therefore only those at zero-energy, and the near-resonant behavior is the same for positive and negative $\epsilon$.\footnote{Resonances are sometimes characterized as zeroes of the Jost function \cite{Watanabe:2025kgw}, and such zeroes may also occur in the $s$-wave case far away from the zero-energy region (we thank Yuki Watanabe for bringing this point to our attention), but to our understanding it is the zero-energy resonances that are principally relevant for unitarization in the $s$-wave case.}

\subsection{The equivalence between the PSS24 regulator and W25 regulator}
\label{sec:sec_equiv_a_matching_RG_ss}
To relate the PSS24 and W25 unitarization prescriptions, we begin by relating their short-distance annihilation amplitudes $\bar f_{s,\ell}$ and $f_\ell^S$, respectively. 
We first express~(\ref{eq:W25_fS}) in terms of the regular solution for the scattering problem when the potential is set to zero outside of $r=a$,  $F_\ell^a$. Since the regular solution to the Schr\"odinger equation is uniquely determined up to an overall normalization factor, we must have $F_\ell^a = (C_\ell^a/C_\ell) F_\ell$. The amplitude is then
\begin{align}
\label{eq:PSS24_W25_fsl_relation}
    f_\ell^S = -\frac{2\mu}{p^2{C_\ell^a}^2} \int_{0}^{\infty} dr F^a_\ell(r) V_S(r)F^a_\ell(r).
\end{align}
Note that in general, the short-ranged potential is a function of two positional arguments $V_S(r,r')$, the arguments of the $F_\ell^a$ above are $r$ and $r'$, and there is an additional integral over $r'$. In the W25 approach, it is assumed the potential is local $V_S(r,r')=V_S(r)\delta(r-r')$. Thus, for small evolution within the matching radius, we have $F_\ell^a(r)\sim C_\ell^a j_\ell(r)$.
The above is therefore approximately the Born amplitude for scattering off the short-ranged potential $V_S(r)$, so we can associate 
\begin{align}
\label{eq:RG_short_distance_association}
    \bar f_{s,\ell}\simeq f_\ell^S.
\end{align}

Returning now to the cross sections~(\ref{eq:PSS24_cross_section_ss}) and~(\ref{eq:W25_cross_section}), we see that if we can associate
\begin{equation}
\label{eq:Z_cotdelta}
    \bar Z_\ell \simeq pC_\ell^2\cot\delta_\ell, 
\end{equation}
the PSS24 and W25 cross sections coincide. Note that the above cannot be an equality since the left-hand side of~(\ref{eq:Z_cotdelta}) depends on the matching radius $a$ and the right-hand side does not; however, as discussed previously, in the following we will show that where the $\bar{Z}_\ell$ correction is large, it is nearly $a$-independent. It is the large, $a$-independent part of $\bar Z_\ell$ that coincides with $p C_\ell^2 \cot\delta_\ell$ up to parametrically suppressed terms.

\subsubsection{Approximate $a$-independent form for $\bar{Z}_\ell$}
We now seek to understand the behavior of the PSS24 regulator $\bar Z_\ell$ in more detail. We first express $\bar Z_\ell$ in terms of the regular and irregular solutions with and without the long-range potential. We have 
\begin{equation}\label{eq:Z_ell_ss_wronskian}
\bar Z_\ell = pC_\ell^2\frac{W_a[c_\ell + is_\ell,G_\ell]}{W_a[c_\ell + is_\ell,F_\ell]}.
\end{equation}
To obtain~(\ref{eq:Z_ell_ss_wronskian}), we have solved for the Sommerfeld factor in the absence of potential outside the matching radius, $Q_\ell^a = p C_\ell/W_a[c_\ell + i s_\ell,F_\ell]$. The result for $Q_\ell^a$ is obtained by matching the regular solution to free-particle propagation at $r=a$ for an incoming unit-normalized plane wave.

Since we assume $p/\mu\ll 1$ and $a\sim 1/\mu$, and that the evolution due to the long range potential is on scales $R  \gg 1/\mu$,
we can expand the regular and irregular solutions in terms of their dominant terms to understand their behavior near $r\sim a$. We perform a power-series expansion around $r=0$ in the limit $p r \ll 1$, and then evaluate the result at $r=a$; provided $p a \ll 1$ (which should be true for non-relativistic momenta and $a\sim 1/\mu$), we expect this expansion to be valid. The leading behavior of the solutions for small $p,r$ are (see Appendix~\ref{app:analyticwavefn})
\begin{align}
\label{eq:FG_a_leading_order}
F_\ell(r)&\simeq C_\ell \frac{(p r)^{\ell+1}}{(2\ell+1)!!} \nonumber \\
G_\ell(r)&\simeq C_\ell^{-1}(2\ell-1)!!(p r)^{-\ell} + g_{\ell,\ell+1} (p r)^{\ell+1}.
\end{align}
Note that while the second term in $G_\ell(r)$ appears parametrically subdominant, the coefficient $g_{\ell,\ell+1}$ is not fixed by the  Frobenius relations at the origin, and must be fixed by the boundary conditions at infinity. In particular, this coefficient $g_{\ell,\ell+1}$ can become very large on resonances, and so it must be included for an accurate description of $G_\ell(r)$, even though we drop subdominant terms whose coefficients (relative to the leading terms) are fully determined by the Schr\"odinger equation at the origin and so are not sensitive to the resonances.

Substituting~(\ref{eq:FG_a_leading_order}) into~(\ref{eq:Z_ell_ss_wronskian}) and expanding the free-particle solutions to leading order in $p a$, the leading behavior of the regulator is
\begin{align}\label{eq:Z_ell_LO}
\bar Z_\ell=
\begin{cases}
pC_0 g_{0,1}+\ldots , & \ell=0 ,\\[6pt]
\dfrac{[(2\ell-1)!!]^2}{\ell}\alpha\mu (pa)^{-2\ell}
+(2\ell+1)!!\,pC_\ell g_{\ell,\ell+1}\ldots , & \ell>0 ,
\end{cases}
\end{align}
where $V_L(r)$ is assumed to have the form $V_L(r) =\frac{\alpha}{r} + \rm{finite \,\, terms}$ at short distances. The sub-leading terms above, denoted by $...$, are suppressed by powers of $\mu a, p a, p/\mu, pR, 1/\mu R$ and $\alpha$, where $R$ is the range of the potential (see Appendix~\ref{app:exp_barZell}), and are therefore small so long as $p a \ll 1$ and $a \sim 1/\mu$. 

Recall that in the expression for the regulated cross section (\ref{eq:PSS24_cross_section_ss}), $\bar{Z}_\ell$ appears only in the denominator term $1 - (i p C_\ell^2 + \bar{Z}_\ell) \bar{f}_{s,\ell}$. Consequently, the outgoing-wave regulator can generally be neglected if $|\bar{Z}_\ell \bar{f}_{s,\ell}| \ll 1$. Assuming the universal low-energy behavior of the annihilation amplitude is $\bar f_{s,\ell}\sim i\frac{\alpha}{\mu}\left(\frac{p}{\mu}\right)^{2\ell}$ for some weak coupling $\alpha$, and $a\sim 1/\mu$, the first term in~(\ref{eq:Z_ell_LO}) for $\ell>0$ scales like $\alpha^2$, and so always contributes negligibly to the regulated cross section. Thus the $g_{\ell,\ell+1}$ term represents the only (potentially) non-negligible piece of the outgoing-wave regulator, and so for the purposes of estimating the regulated cross section, we can replace $\bar{Z}_\ell \rightarrow p(2\ell+1)!! C_\ell g_{\ell,\ell+1}$, and can focus on the case where $g_{\ell,\ell+1}$ is large (where this is a poor approximation and/or where $g_{\ell,\ell+1}$ is $\mathcal{O}(1)$, the effect of the outgoing-wave regulator is negligible). In particular, note that this prescription replaces a nominally $a$-dependent quantity with an $a$-independent quantity: the leading $a$-dependence of $\bar{Z}_\ell$ is encoded in the first term of (\ref{eq:Z_ell_LO}), and as we have seen, this term is always small for $a \sim 1/\mu$ when combined with $\bar f_{s,\ell}$.

\subsubsection{Behavior of the regular and irregular solutions near a resonance}

Now that we know $\bar{Z}_\ell$ is controlled by $g_{\ell,\ell+1}$ whenever it is relevant to the regulated cross section, we wish to understand the circumstances under which $g_{\ell,\ell+1}$ is parametrically large, and, as a result, $\bar Z_\ell$ is dominated by a large, $a$-independent constant. 

We begin by expressing the regular and irregular solutions in~(\ref{eq:Z_ell_ss_wronskian}) in terms of the regular and irregular solutions when the long-range potential is set to zero outside the matching radius. In addition to the regular solution $F_\ell^a$ we have already discussed, we define the irregular $G_\ell^a$ solution in this case to have the corresponding boundary conditions~(\ref{eq:shortrangebcs}-\ref{eq:longrangebcs}) with $\delta_\ell \rightarrow \delta_\ell^a$ and $C_\ell \rightarrow C_\ell^a$. For $r<a$, $G_\ell^a(r)$ and $F_\ell^a(r)$ can be written as linear combinations of the full solutions $F_\ell(r)$ and $G_\ell(r)$ (since they solve the same Schr\"{o}dinger equation).  
Since $G_\ell^a$ is irregular, we must have $G_\ell^a = (C_\ell/C_\ell^a)G_\ell + \beta F_\ell$, for some constant $\beta$. The coefficient $\beta$ is obtained by matching $G_\ell^a + iF_\ell^a$ onto a purely outgoing wave at $r=a$. The result of this matching is 
\begin{align} 
G_\ell(r) & = \frac{C^a_\ell}{C_\ell} (G_\ell^a(r) + i F_\ell^a(r)) + \frac{C_\ell}{C_\ell^a} \frac{\alpha_{G_\ell}(a)}{\alpha_{F_\ell}(a)} F_\ell^a(r),  \quad F_\ell(r) = (C_\ell/C_\ell^a) F_\ell^a(r).\label{eq:matching}
\end{align}

At $r=a$, $G^a_\ell(r) + i F_\ell^a(r)$ must match onto a purely outgoing free-particle solution (i.e.~proportional to $c_\ell(p r) + i s_\ell(p r)$). We do not expect large enhancements in the normalization of  $G^a_\ell(r) + i F_\ell^a(r)$ provided the short-range physics is weakly coupled, and so the outgoing part of $G_\ell(r)$ is suppressed at $r=a$ when the Sommerfeld factor $C_\ell$ is parametrically large (as happens close to a resonance). Thus, we see that in this regime,
the irregular solution $G_\ell(r)$ behaves (for $r \sim a$) like the short-range regular solution scaled by a factor $\frac{C_\ell}{C_\ell^a} \frac{\alpha_{G_\ell}(a)}{\alpha_{F_\ell}(a)}$, or $F_\ell(r)$ rescaled by a factor $\frac{\alpha_{G_\ell}(a)}{\alpha_{F_\ell}(a)}$. This suggests that where there is a large Sommerfeld factor $C_\ell$, we will also be in the regime where the $g_{\ell,\ell+1}$ term is large in (\ref{eq:FG_a_leading_order}), and we can identify $g_{\ell,\ell+1} (2\ell+1)!!/C_\ell \simeq \frac{\alpha_{G_\ell}(a)}{\alpha_{F_\ell}(a)}$ as the rescaling of the irregular solution relative to the regular solution (at $r\sim a$). This result is consistent with the intuition that the $g_{\ell,\ell+1}$ term controls the weight of the regular solution in the irregular solution. 

There is another qualitative way to understand why the irregular, purely outgoing solution increasingly resembles the regular solution near resonance points. The classic Sommerfeld resonances are associated with the onset of bound states in the spectrum, and bound states correspond to negative-energy (imaginary-momentum) solutions whose wavefunctions are simultaneously exponentially falling at large $r$ ($\propto e^{-\kappa r}$) and regular as $r\rightarrow 0$. We can replace $p=i\kappa$ in the threshold bound-state, and the resulting wavefunction will be regular and appear purely outgoing at large $r$ ($\propto e^{i p r}$). In the limit $\kappa, p \rightarrow 0$, which corresponds to the resonance point when the bound state first enters the spectrum, these states coincide, and the near-zero-energy, purely outgoing scattering state should approach the regular solution.

\begin{figure}[t]
    \centering
    \includegraphics[width=\textwidth]{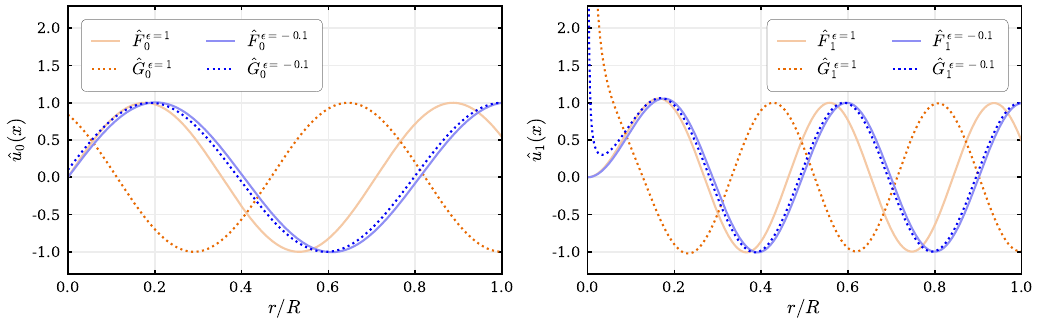}
    \caption{Plots of the regular and irregular solutions for the finite spherical well potential $V_L(r)=-\frac{p_V^2}{2\mu}\theta(R-r)$ for $\ell=0$ (left) and $\ell=1$ (right) near (blue) and far (orange) from resonance. The depth of the potential is chosen such that $p_V=(5\pi +\epsilon)/2^{1-\ell} R$, where $\epsilon=0$ corresponds to resonances for both $\ell=0$ and $\ell=1$. The regular and irregular solutions $\hat F_\ell$ and $\hat G_\ell$ are normalized to their peak value nearest the range $R$ of the spherical well. Far from resonance, for $\epsilon=1$, there is little correlation between the solutions within $r<R$. Near-resonance, for $\epsilon=-0.1$, sufficiently far from the origin, the regular and irregular solutions are seen to be approximately scaled relative to each other. 
    }
    \label{fig:FG_ratio}
\end{figure}

In order to relate the proportionality factor $\alpha_{G_\ell}(a)/\alpha_{F_\ell}(a)$ to the $S$-matrix, we consider the behavior of the solutions near the range of the potential $r \gtrsim R$ (e.g.~for a Yukawa potential, $R\sim 1/m_\phi$ where $m_\phi$ is the mass of the mediator). We assume negligible evolution due to the long range potential beyond $r\sim R$, so we can continue the asymptotic free-particle solutions into $R$ from infinity:
\begin{align}
\label{eq:FG_at_V_range}
    F_\ell( R)&\simeq s_\ell(\bar p )\cos\delta_\ell+c_\ell(\bar p)\sin\delta_\ell \\
    G_\ell( R)&\simeq c_\ell(\bar p)\cos\delta_\ell-s_\ell(\bar p)\sin\delta_\ell,
\end{align}
where $\bar p \equiv pR$. We observe that in the limit $\bar{p} \rightarrow 0$ with $R$ held fixed, the $s_\ell(\bar{p})$ terms become subdominant near $r=R$ and we expect both $F_\ell(r)$ and $G_\ell(r)$ to behave as $c_\ell(p r)$ for $r \sim R$, with coefficients fixed by $\cos\delta_\ell$ and $\sin\delta_\ell$ respectively. Note that it is for these momenta, much smaller than the inverse range of the potential, where we generally observe the effects of resonant Sommerfeld enhancement, see e.g.~the discussion in Ref.~\cite{ArkaniHamed:2008qn}. More precisely, we can consider the log-derivative of the solutions near the range of the potential. We expand the trigonometric factors involving the phase-shift according to~(\ref{eq:ERE_ss}) and the free particle solutions to leading-order in $p$ to obtain
\begin{align}
\label{eq:log_derivative_range}
    F_\ell^{-1}(R)F_\ell'(R)&=c_\ell^{-1}(\bar p)c_\ell'(\bar p)\Big(1 + \mathcal O(\bar p^{2\ell+1}\cot\delta_\ell)\Big) \\
    G_\ell^{-1}(R)G_\ell'(R)&=c_\ell^{-1}(\bar p)c_\ell'(\bar p)\Big(1 + \mathcal O(\bar p^{2\ell+1}\tan\delta_\ell)\Big).
\end{align}
The quantities inside the $\mathcal O$ brackets above are both small and negligible for small $p$ near a resonance (and on a resonance for $\ell>0$). That the first line is small is clear from the effective-range expansion~(\ref{eq:ERE_ss}). The term on the second line scales like
\begin{align}
\label{eq:ERE_expand_tan}
    p^{2\ell+1}\tan\delta_\ell&= \frac{p^{4\ell+2}}{p^{2\ell+1}\cot\delta_\ell}\nonumber \\
    &= \frac{p^{4\ell+2}}{\epsilon^{2\ell+1} + \frac{1}{2r_\ell^{2\ell-1}}p^2 + \mathcal{O}(p^4)},
\end{align}
and is therefore small for $\ell=0$ near-resonance for sufficiently small $p$ and for $\ell>0$ for $p\ll 1/R$ due to the $p^{4\ell+2}$ scaling in the numerator.
In these cases, the log-derivatives of the solutions therefore approximately coincide near the range of the potential near a resonance, i.e. $G_\ell \propto F_\ell$ near $r\sim R$ when the scattering system is close to a resonance. The log-derivatives of the regular and irregular solutions can never exactly coincide, as the small, crucially different, corrections to~(\ref{eq:log_derivative_range}) yield solutions that asymptote to sine waves that are $\pi/2$ out-of-phase at infinity. Note for $\ell=0$, precisely on-resonance ($\epsilon=0$), the above quantity is not necessarily small, and the logarithmic derivatives of $F_\ell$ and $G_\ell$ do not necessarily coincide near the range of the potential. In this case, however,  the outgoing-wave regulator is small and negligible as well, and the $p C_0^2 \bar f_{s,0}$ term in~(\ref{eq:PSS24_cross_section_ss}) supplies the large, unitarity preserving corrections. To see this, note that since the Sommerfeld factor scales like $C_0\sim 1/p$ on-resonance, the W25 regulator is a constant on-resonance ($pC_0^2\cot\delta_0=\rm{const.}$), and will therefore be small as compared with $pC_0^2$ for sufficiently small $p$.  

At $r\sim R$, when the log-derivatives approximately coincide, the proportionality factor between $G_\ell(r)$ and $F_\ell(r)$ is $\simeq \cot\delta_\ell$, since both solutions are dominated by their respective $c_\ell$ term in~(\ref{eq:FG_at_V_range}). As we have argued above, when $g_{\ell,\ell+1}$ is large (which is automatic for a sufficiently large Sommerfeld enhancement factor $C_\ell$), $G_\ell(r)$ is approximately proportional to $F_\ell(r)$ already at $r=a$. Since the two solutions $F_\ell(r)$ and $G_\ell(r)$ evolve under the same Schr\"{o}dinger equation between $r=a$ and $r=R$, the proportionality factors at $r=a$ and $r=R$ can be identified. We thus obtain
\begin{align}\cot \delta_\ell  \simeq \frac{\alpha_{G_\ell}(a)}{\alpha_{F_\ell}(a)} =  \frac{W_a[c_\ell + is_\ell, G_\ell]}{W_a[c_\ell + is_\ell, F_\ell]}\simeq g_{\ell,\ell+1} (2\ell+1)!!/C_\ell. \label{eq:Wronskian_ratio_identification} \end{align}
As a consequence, $\bar{Z}_\ell$ can be written in the simple, $a$-independent form $\bar{Z}_\ell \rightarrow p C_\ell^2 \cot \delta_\ell$. So far these arguments have been qualitative; in the following section, we provide a more quantitative argument for when the above identification can be made, and for its validity for $\ell=0$ on-resonance, when the logarithmic derivatives of $G_\ell$ and $F_\ell$ at the range of the potential do not approximately coincide. Figure~\ref{fig:FG_ratio} shows an example of the proportionality of $G_\ell(r)$ and $F_\ell(r)$ in the resonant regime, for the simple example of the spherical well potential $V_L(r)=-\frac{p_V^2}{2\mu}\theta(R-r)$.

\subsubsection{Validating this approximation with the series expansions of $F_\ell$ and $G_\ell$}
\label{sec:frobenius_ss}
To get a better analytic handle on the validity of~(\ref{eq:Wronskian_ratio_identification}), we consider the leading-order behavior of the regular and irregular solutions for $p\ll 1/R,\mu$ near the range of the potential. We match the expansions for $F_\ell$ and $G_\ell$ to the free-particle solutions~(\ref{eq:FG_at_V_range}) near the range of the potential to understand the dependence of the short-distance behavior of the solutions on the $S$-matrix. For small $p$, we expand $F_\ell$ and $G_\ell$ according to
\begin{align}
\label{eq:FG_small_p_expansion}F_\ell(r)&=p^{\ell+1}C_\ell \big(F^{(p=0)}_\ell(r) + \mathcal{O}(p^2)\big) \\
    G_\ell(r)&=p^{-\ell}C_\ell^{-1} \big(G^{(p=0)}_{\ell,-}(r) + \mathcal{O}(p^2)\big) + (2\ell+1)!!g_{\ell,\ell+1}p^{\ell+1} G_{\ell,+}^{(p=0)}(r),
\end{align}
where $F_{\ell}^{(p=0)}$ is the zero-energy regular solution to the Schr\"odinger equation with boundary condition $F^{(p=0)}_\ell(r\rightarrow 0)\rightarrow r^{\ell+1}/(2\ell+1)!!$ and $G^{(p=0)}_{\ell,-}(r)$ and $G^{(p=0)}_{\ell,+}(r)$ contain powers of $r$ less than $r^{\ell+1}$ and greater or equal to $r^{\ell+1}$, respectively. $G_\ell(r)$ also generically contains a logarithmic term proportional to the regular solution, which, for our purposes, we can neglect (see the discussion in Appendix~\ref{app:analyticwavefn} for details). In short, this term scales at least like $p^{\ell+1}\log pr$ for small $p$, so it will be negligible when matching the dominant contributions of the series expansion of $G_\ell(r)$ to its expansion in terms of the free-particle solutions near the range of the potential.

To obtain an estimate for the Sommerfeld factor $C_\ell$ and the $g_{\ell,\ell+1}$ coefficient, we match the expression for $F_\ell$ and $G_\ell$ in terms of free particle solutions to the zero-$p$ expansions above and expand the trigonometric factors according to~(\ref{eq:ERE_ss}).
We find
\begin{align}
\label{eq:C_approx_frobenius}
    C_\ell\simeq F_\ell^{(p=0)}(R)^{-1}\frac{(2\ell-1)!!(2\ell+1)!!}{\sqrt{\bar p^{4\ell+2} + \big(\bar \epsilon^{2\ell+1}+ \frac{\bar p^2}{2( r_\ell/R)^{2\ell-1}}\big)^2}}\Big(1 + \mathcal{O}(\bar p^2, \bar p^{2\ell+1}\cot \delta_\ell,\bar p^{-2\ell+3}\tan \delta_\ell\Big),
\end{align}
for the Sommerfeld factor. That the second term in the $\mathcal{O}$ brackets above is small follows from~(\ref{eq:ERE_expand_tan}), only now the scaling in the numerator is $p^4$ rather than $p^{4\ell+2}$. We note that we observe a similar behavior of the Sommerfeld factor near-resonance as is obtained via the WKB method in Ref.~\cite{Beneke:2024iev}.
Further, assuming $\bar p=pR,\bar\epsilon=\epsilon R\ll1$, we conclude the $g_{\ell,\ell+1}$ term must dominate the expansion~(\ref{eq:FG_small_p_expansion}), so that
\begin{equation}
g_{\ell,\ell+1}\simeq (2\ell-1)!!{G_{\ell,+}^{(p=0)}}(R)^{-1}\bar p^{-2\ell-1}\frac{\bar\epsilon^{2\ell+1} + \frac{\bar p^2}{2(r_\ell/R)^{2\ell-1}}}{\sqrt{\bar p^{4\ell+2} + \Big(\bar\epsilon^{2\ell+1} + \frac{\bar p^2}{2(r_\ell/R)^{2\ell-1}}\Big)^2}}\Big(1 + \mathcal{O}(\bar p^2, \bar p^{2\ell+1}\cot\delta_\ell)\Big).
\end{equation}
The coefficient of $r^{\ell+1}$ therefore dominates the expansion of the irregular solution at the range of the potential. We see here that precisely on a resonance, in the $s$-wave case the potentially large part of the outgoing-wave regulator $\bar Z_0\simeq pC_0g_{0,1}\simeq \rm{const.}$, and so is negligible as compared to the $pC_0^2\sim 1/p$ term, consistent with our earlier arguments. 

It remains to be understood the behavior of the functions $F_\ell^{(p=0)}(R)$ and ${G_{\ell,+}^{(p=0)}}(R)$, which depend only on the resonance parameters $\alpha$ and $\mu R$. Since the coefficients of the irregular solution for powers of $r^k$ greater than $r^\ell$ are, at zeroth-order, governed by the same recursion relations at the regular solution (since the $g_{\ell,\ell+1}$ term dominates), the polynomial ${G_{\ell,+}^{(p=0)}}$ is approximately equal to $F_\ell^{(p=0)}$ near resonance. Another way to see this is by considering again the logarithmic derivative of $F_\ell$ and $G_\ell$ near the range of the potential when $\bar p,\bar\epsilon\ll1$. That the logarithmic derivatives~(\ref{eq:log_derivative_range}) are approximately equal at the range of the potential implies $G_{\ell,+}^{(p=0)}(R)\simeq F_{\ell}^{(p=0)}(R)$ up to terms parametrically suppressed in velocity. The expansions above therefore yield $\cot\delta_\ell\simeq (2\ell+1)!!C_\ell^{-1} g_{\ell,\ell+1}$ near resonance, and the leading-order (and only relevant) terms in the Wronskian expansion~(\ref{eq:Z_ell_LO}) can be written:
\begin{align}
\label{eq:Zbarell_equivalence}
\bar Z_\ell \simeq (2\ell+1)!!pC_\ell g_{\ell,\ell+1}\simeq p C_\ell^2\cot\delta_\ell.
\end{align}
We therefore conclude that the regulated $S$-matrix obtained via the PSS24 and W25 procedures approximately coincide when the corrections that preserve unitarity are large.

We note that far from resonance, when $\epsilon R\gg 1$, both the regulators, when combined with the short distance annihilation amplitude $\bar f_{s,\ell}\sim i\frac{\alpha}{\mu}(\frac{p}{\mu})^{2\ell}$, are negligible corrections to the cross section. To see this, note that far from resonance, the first term in~(\ref{eq:FG_a_leading_order}) in $G_\ell$ dominates its expansion at $r\sim a$. $G_\ell$ therefore behaves like $c_\ell$ near the origin, and the numerator in~(\ref{eq:Z_ell_ss_wronskian}) receives no enhancement. Thus the regulator goes like $\bar Z_\ell\bar f_{s,\ell}\sim \alpha^2$ for small $p$. Further, for large $p$, $C_\ell\sim 1$, so the regulator is negligible. The same is true for the W25 regulator, noting~(\ref{eq:ERE_ss}) and (\ref{eq:C_approx_frobenius}).

\subsection{Relating the FP25 approach to the PSS24 approach}
We now demonstrate when and how the regulated $S$-matrix obtained via the FP25 approach coincides with that of the PSS24 approach. Our approach builds on Appendix C of PSS24, but avoids some approximations made in that work. As discussed previously, the outgoing-wave regulator in the FP25 approach may contain UV divergences; matching this approach to that of PSS24 clarifies how these UV divergences are captured in the PSS24 analysis (and the W25 analysis, which is similar to PSS24 in this regard).

To relate the PSS24 and FP25 results, we will first seek to estimate the relation between the short-distance amplitude $\bar{f}_{s,\ell}$ and the unregulated amplitude defined in the FP25 method. We will find that the bare annihilation amplitude can be identified at leading order with the amplitude obtained via the DWBA, 
with the unperturbed wavefunctions given by the regular solution $F_\ell^a$ for the purely short-range scattering problem, i.e.~when the long range potential is set to zero outside the matching radius. 

In the FP25 formalism, the unregulated annihilation amplitude is given by (\ref{eq:FP_unregulated_amplitude}). In the case studied by PSS24 and W25, where the absorptive interactions encoded by $v_\ell(r)$ are short-range, we can approximate this integral as zero outside $r=a$. Furthermore, in the region $r < a$, we have a simple proportionality relationship between $F_\ell(r)$ and the regular wavefunction in the presence of only the short-range part of the potential, $F^a_\ell(r)$, as derived in (\ref{eq:matching}). Under this short-range assumption we can write:
\begin{align} M_\ell^{\text{unreg.}} & \simeq C_\ell \left[ \frac{1}{C_\ell^a} \sqrt{\frac{1}{p}} \int^a_0 dr r F_\ell^a(r) v_\ell(r) \right], \end{align}
where the term in square brackets is associated purely with short-range physics (which can ideally be computed from a weakly-coupled relativistic UV theory without the need for resummation). Noting~(\ref{eq:PSS24_W25_fsl_relation}),~(\ref{eq:RG_short_distance_association}), and that the short-ranged potential is given by $2\mu V_S(r,r') = -\eta_\ell^2 r r' v_\ell(r)v^*_\ell(r')$, the Born approximation for $\bar{f}_{s,\ell}$ gives the leading-order identification as:
\begin{align}
    \eta_\ell^2 |M_\ell^{\rm unreg.}|^2 \simeq p C_\ell^2 \bar f_{s,\ell}, 
\end{align}
with each side encoding the Sommerfeld-enhanced but unregulated cross section.
With this association, we note that
if we can also associate
\begin{align}
\label{eq:self_energy_regulator}
    \frac{W_\ell}{|M^{\rm unreg.}_\ell|^2}\simeq \frac{W_a[c_\ell+is_\ell,G_\ell]}{W_a[c_\ell+is_\ell,F_\ell]} = \bar{Z}_\ell/(p C_\ell^2),
\end{align}
when the corrections are large, the PSS24 $S$-matrix~(\ref{eq:PSS-Smatrix}) and the FP25 $S$-matrix~(\ref{eq:FP-Smatrix}) will approximately coincide. Motivated by this estimate, we will now compute $\frac{W_\ell}{|M^{\rm unreg.}_\ell|^2}$ (although we will demonstrate that this guess is only correct to leading order and the true matching is slightly more subtle). 

To relate the FP25 quantities above, we expand their definitions~(\ref{eq:FP_unregulated_amplitude}),~(\ref{eq:FP_Wl_definition}) in terms of the scattering solutions when the potential is cut-off at $r=a$. Under the same assumption that integrals involving $v_\ell(r)$ factors  can be truncated at $r=a$ (this encodes the assumption in PSS24 that absorptive interactions are purely short-range), and again using (\ref{eq:matching}) in the $0 < r < a$ region,  the ratio of regulators in the FP25 approach is 
\begin{align} \frac{W_\ell}{|M^{\rm unreg.}_\ell|^2} & = \frac{\int dr r \int dr^\prime r^\prime F_\ell(r_<) G_\ell(r_>) v_\ell(r) v_\ell(r^\prime)}{\int dr r \int dr^\prime r^\prime F_\ell(r_<) F_\ell(r_>) v_\ell(r) v_\ell(r^\prime)} \nonumber \\
& = \frac{1}{(C_\ell/C_\ell^a)} \frac{1}{\int_0^a dr r \int dr^\prime r^\prime F^a_\ell(r) F^a_\ell(r') v_\ell(r) v_\ell(r^\prime)} \nonumber \\ 
& \times \int_0^a dr r \int_0^a dr^\prime r^\prime F^a_\ell(r_<) ((C_\ell^a/C_\ell) (G_\ell^a(r_>) + i F_\ell^a(r_>)) + \frac{C_\ell}{C_\ell^a} \frac{\alpha_{G_\ell}(a)}{\alpha_{F_\ell}(a)} F^a_\ell(r_>)) v_\ell(r) v_\ell(r^\prime) \nonumber \\
& =\frac{1}{(C_\ell/C_\ell^a)}\left( \frac{C_\ell}{C_\ell^a} \frac{\alpha_{G_\ell}(a)}{\alpha_{F_\ell}(a)} +  (C_\ell^a/C_\ell) \frac{ \Lambda_1(a) p/\eta_\ell^2}{\Lambda(a) p^2 (C^a_\ell)^2 / \eta_\ell^2} \right) \nonumber \\
&=  \frac{\alpha_{G_\ell}(a)}{\alpha_{F_\ell}(a)} +   \frac{\Lambda_1(a)}{p C_\ell^2 \Lambda(a)  }\label{eq:FPmatching-split}
\end{align}
where we have defined the two integrals, both depending solely on short-range physics:
\begin{align} \Lambda (a) & = \frac{\eta_\ell^2}{ p^2 (C_\ell^a)^2} \left|\int^a_0 r dr v_\ell(r)  F_{\ell}^{a}(r) \right|^2 = \frac{\eta_\ell^2}{p C_\ell^2} |M_\ell^\text{unreg.}|^2, \nonumber \\
\Lambda_1(a) &\equiv \frac{\eta_\ell^2}{p} \int^a_0 r dr \int^a_0 r^\prime dr^\prime v_\ell(r) v_\ell^*(r^\prime) F^{a*}_\ell(r_<) (G^a_\ell(r_>) + i F^a_\ell(r_>)).
\end{align}
 Note that $\Lambda_1(a)$ can diverge due to the integral of the irregular solution, as discussed previously, but we can now see that this is a purely UV divergence that has nothing to do with the IR enhancement from the long-range potential. Accordingly, we expect that in the PSS24 formalism, it will be captured by the amplitude describing the short-range physics. 
 Rewriting the FP25 $S$-matrix (\ref{eq:FP-Smatrix}) in terms of the above quantities,
 we have
\begin{align} 
S_\ell & = S_{0,\ell} \left[1 + \frac{2i \eta_\ell^2 |M_\ell^\text{unreg.}|^2}{1 - \eta_\ell^2 |M_\ell^\text{unreg.}|^2 (W_\ell/|M_\ell^\text{unreg.}|^2 + i)  } \right] \nonumber \\
& = S_{0,\ell} \left[1 + \frac{2i p C_\ell^2 \Lambda(a)}{1 -   \Lambda_1(a) - \Lambda(a) \left( i p C_\ell^2 +  p C_\ell^2 \frac{\alpha_{G_\ell}(a)}{\alpha_{F_\ell}(a)}\right)   } \right]  \nonumber \\
& = S_{0,\ell} \left[1 + \frac{2i p C_\ell^2 }{(1 -  \Lambda_1(a))/\Lambda(a)  - \left(i p  C_\ell^2  + p  
C_\ell^2 \frac{\alpha_{G_\ell}(a)}{\alpha_{F_\ell}(a)} \right) } \right]
\label{eq:FPmatching-Smatrix} \end{align}
Comparing to (\ref{eq:PSS-Smatrix}), we see that the expressions will match precisely if we can relate the short-range-only amplitudes via:
\begin{align} 
 \bar{f}_{s,\ell} & = \frac{\Lambda(a)}{1 - \Lambda_1(a)},\label{eq:FPmatching}
\end{align}
and the long-range outgoing-wave regulator term by:
\begin{align} \bar{Z}_\ell = p C_\ell^2 \frac{\alpha_{G_\ell}(a)}{\alpha_{F_\ell}(a)}. \end{align}
This latter expression precisely matches our previously derived result for $\bar{Z}_\ell$ (\ref{eq:Z_ell_ss_wronskian}). 

Since both $\Lambda(a)$ and $\Lambda_1(a)$ include factors of the short-range absorptive potential encoded in the $v_\ell(r)$ functions, and its accompanying coupling $\eta_\ell^2$, they are higher order in the coupling governing this process. The lowest-order matching for the short-range physics thus corresponds to $\bar{f}_{s,\ell} =\Lambda(a)$, with  corrections that are higher-order in the coupling from a (possibly divergent) $\Lambda_1(a)$. The leading-order matching corresponds to what we would expect from equating the annihilation cross section obtained from the leading-order DWBA (in $\Lambda(a)$), with the imaginary part of the (Sommerfeld-enhanced) short-range scattering amplitude $(C^a_\ell)^2 \bar{f}_{s,\ell}$. This is intuitively reasonable and in agreement with the optical theorem. At higher order, we interpret divergent contributions to $\Lambda_1(a)$ as representing the familiar UV divergences that can appear in higher-order (loop) diagrams when attempting to relate bare and renormalized quantities; as shown here, they are distinct and can be separated from the IR enhancements associated with the long-range potential. The form of the matching in (\ref{eq:FPmatching}) is reminiscent of the renormalization of the propagator by summing up an infinite series of 1PI insertions, although we do not aim here to derive this relationship independently from first principles.

The regulators for the FP25 and PSS24 method therefore coincide exactly, whether the system is near a zero- or finite-energy resonance or not, provided the short-range amplitudes are correctly matched.  In this case the regulated $S$-matrices also coincide, ensuring that the corrected cross sections for both annihilation and elastic scattering will also match.

The $S$-matrix as we have described it here, projected onto the non-relativistic two-body states described by QM, does not describe the {\it exclusive} annihilation cross sections for different final states, in the more general multi-channel case (it only has access to the inclusive annihilation cross section because this can be written in terms of the apparent lack of unitarity in the QM treatment, due to probability loss into annihilation). However, FP25 computes results for the regulated exclusive annihilation amplitudes / cross sections; in Appendix~\ref{app:multichannel} we work out the analogue of (\ref{eq:FPmatching-split}) for the multi-channel case (which is conceptually identical to the single-channel case), and confirm this reproduces the leading-order PSS result for exclusive cross sections given in Ref.~\cite{Parikh:2024mwa}.

In Ref.~\cite{Flores:2025uoh}, there is a discussion of the necessity of renormalization in the case of contact interactions, focusing for simplicity on the case of a single inelastic channel. In particular, \texttt{v2} of that work considers renormalization in the context of a delta-function potential with $\ell=0$, as studied in Ref.~\cite{Blum:2016nrz} (this case was explicitly cross-checked against the PSS24 formalism in Ref.~\cite{Parikh:2024mwa}). Their analysis finds that for $s$-wave renormalizable contact interactions, there is a UV divergence (which is momentum-independent in the $s$-wave case) that can be absorbed into the complex coupling of the contact interaction, and separately a finite momentum-dependent part that contributes to the regularization.  (\ref{eq:FPmatching-split}) demonstrates the explicit separation of these two contributions, for the more general case where higher partial waves are allowed and the short-range physics need not be described by a delta-function potential. 

For comparison, note that \texttt{v1} of Ref.~\cite{Flores:2025uoh} neglected the non-divergent but momentum-dependent term in their renormalization procedure, and consequently obtained an overly simplified expression for the regulated cross section (in Eq.~4.23(b) of that version), corresponding to setting the $\bar{Z}_\ell$ regulator (or in their notation, the $\mathbb{W}_\ell$ term) to zero. This result resembles that obtained in Ref.~\cite{Flores:2024sfy} under certain simplifying assumptions for the high-momentum behavior of the amplitudes (see that work for details). Based on our comparison with W25 allowing us to express $\bar{Z}_\ell$ in terms of the elastic-scattering phase shift near resonances, we do not generally expect this regulator term to be zero for any long-range potential that exhibits resonant behavior.

\subsection{Example: the Hulth\'{e}n potential}
In this section, we validate the identification~(\ref{eq:Zbarell_equivalence}) by considering the example of scattering via the Hulth\'{e}n potential. In particular, we compute the outgoing-wave regulator for the PSS24 and W25 approaches, and show explicitly that the regulators coincide near a resonance. 

\subsubsection{General results}

The Hulth\'{e}n potential is
\begin{align}
    V_L^H(r) = -\frac{\alpha m_\phi e^{-m_\phi r}}{1-e^{-m_\phi r}},
\end{align}
which is exactly solvable in the $s$-wave case. To solve the Schr\"odinger equation for $p$-wave scattering or higher, one can solve the Schr\"odinger equation numerically or modify the centrifugal term according to
\begin{align}
    \frac{\ell(\ell+1)}{r^2}\rightarrow \frac{\ell(\ell+1)m_\phi^2e^{-m_\phi r}}{(1-e^{-m_\phi r})^2},
\end{align}
which yields a modified Schr\"odinger equation whose solutions can be written in terms of hypergeometric functions.
The above modification has the desired properties that it matches the behavior of the usual centrifugal term near the origin and vanishes at infinity. The regular and irregular solutions near the origin behave as polynomials with leading powers $r^{\ell+1}$ and $r^{-\ell}$, respectively, while at infinity they asymptote to sine-waves. However, the exponential decay of the modified centrifugal term is much faster than the polynomial decay of $1/r^2$. As a result, the phase shift and Sommerfeld factor near a resonance do not have the correct $p$-dependence~\cite{Kamada:2023iol,Cassel:2009wt}. 

In particular, the exponential decay of the modified centrifugal term yields a Sommerfeld factor $C_\ell$ which scales like $p^{-1}$ for all $\ell$ on-resonance (so that the cross section enhancement scales like $p^{-2}$), whereas the $1/r^2$ centrifugal term yields a Sommerfeld factor $C_\ell$ that goes like $p^{-2}$ (and a cross section enhancement scaling like $p^{-4}$) for $\ell>0$ on a resonance. We therefore caution that the Hulth\'{e}n solution should be used with care when the goal is specifically to study resonant behavior, and we restrict our discussion to $s$-wave scattering where the solution is exact.

The Schr\"odinger equation with the Hulth\'{e}n potential is 
\begin{align}
    \bigg(-\frac{d^2}{dr^2}  - 2\mu \frac{\alpha m_\phi e^{-m_\phi r}}{1-e^{-m_\phi r}} - p^2\bigg)u_0=0,
\end{align}
where $u_0$ denotes the $s$-wave scattering solution. 
Under the change of variables $y=e^{-\delta x}$, where $x = 2\mu\alpha r, \hat p = p/2\mu\alpha, \delta = m_\phi/2\alpha\mu$, the Schr\"odinger equation becomes
\begin{align}
    \bigg(y^2\partial^2_y + y\partial_y  + \frac{\delta^{-1} y}{1-y} + \hat p^2/\delta^2\bigg)u_0=0,
\end{align}
whose solution can now be readily obtained in terms of hypergeometric functions after making the ansatz $u_0(y) = y^{-i\hat p/\delta}x(y)$. The function $x(y)$ satisfies
\begin{align}
    y(1-y)x'' + \big[1 -2i\hat p/\delta + (2i\hat p/\delta -1)y\big]x' + \delta^{-1}x=0,
\end{align}
which is one form of the hypergeometric equation. The solutions are 
\begin{align}\label{hulthen_soln}
    u_0 &= c_1e^{-ipr} {}_2F_1\big(\alpha^-, \alpha^+, 1+\alpha^-+\alpha^+; e^{-m_\phi r}\big) \nonumber \\ 
    &\phantom{==}+c_2 e^{ipr} {}_2F_1\big(- \alpha^-,  - \alpha^+, 1-\alpha^--\alpha^+; e^{-m_\phi r}\big),
\end{align}
where $\alpha_\pm = \frac{i\hat p}{\delta } \pm\frac{\sqrt{\delta-\hat p^2}}{\delta}$. To obtain the purely outgoing solution, we set $c_1=0$ and $|c_2|=1$ (note ${}_2F_1(a,b,c,0)=1$). To get a phase of $0$ on the second term in the $r\rightarrow 0$ limit, we must take 
\begin{align}
c_2 = \exp\bigg[-i\arg \frac{\Gamma(1-\alpha^+)\Gamma(1 - \alpha^-)}{\Gamma(1-\alpha^+ - \alpha^-)}\bigg]
\end{align}
where we have used 
\begin{align}
\lim_{r\rightarrow 0}\:\: &e^{ipr} {}_2F_1\big( - \alpha^-,  - \alpha^+, 1-\alpha^--\alpha^+; e^{-m_\phi r}\big)  =\frac{\Gamma(1-\alpha^+-\alpha^-)}{\Gamma(1-\alpha^-)\Gamma(1-\alpha^+)}.
\end{align}
The Sommerfeld enhancement factor is thus
\begin{align}\label{eq:hulthen_sommfac}
    C_0 &= \left|\frac{\Gamma(1-\alpha^-)\Gamma(1-\alpha^+)|}{\Gamma(1 + 2i\hat p/\delta)} \right| \nonumber \\
    &=\frac{|\Gamma(1-\alpha^-)||\Gamma(1-\alpha^+)|}{\sqrt{\frac{\pi \hat p/\delta}{\sinh(\pi\hat p/\delta)}}},
\end{align}
the phase-shift is
\begin{align}
    \delta_0 = \arg \bigg\{\frac{\Gamma(1-\alpha^-)\Gamma(1-\alpha^+)}{\Gamma(1 - \alpha^+ -\alpha^-)}\bigg\},
\end{align}
and the purely outgoing solution is
\begin{align}\label{eq:scattering_solutions}
    G_0 + iF_0 = e^{ipr + i\delta_0} {}_2F_1(-\alpha^+,  - \alpha^-, 1-\alpha^+ - \alpha^-, e^{-m_\phi r}).
\end{align}

To evaluate $\bar Z_0$, we are interested in the expansions of the regular and irregular solutions near the origin. Due to the behavior of the hypergeometric function ${}_2F_1(a,b;c;z)$ at $z=1$, it is difficult to analyze the solutions as a power series at $r=0$. To see the small-$r$ behavior of the solutions, we apply an Euler's-type transformation 
\begin{align}\label{eq:gauss_type_transform}
{}_2F_1(a,b;c;z)
&=
\frac{\Gamma(c)\,\Gamma(c-a-b)}{\Gamma(c-a)\,\Gamma(c-b)}
\,{}_2F_1\!\left(a,b;\,a+b+1-c;\,1-z\right)  \\
&\phantom{===}
+\frac{\Gamma(c)\,\Gamma(a+b-c)}{\Gamma(a)\,\Gamma(b)}
\,(1-z)^{c-a-b}\,
{}_2F_1\!\left(c-a,c-b;\,1+c-a-b;\,1-z\right) \nonumber
\end{align}
since the hypergeometric function has a regular Laurent series.
We cannot apply this formula directly, however, since for our solutions, $c-a-b$ is a positive integer (namely, 1), and the hypergeometric series in the first term ${}_2F_1(a,b,c,z)=\sum_n\frac{(a)_n(b)_n}{(c)_n}z^n$, where $(a)_n=\Gamma(a+n)/\Gamma(a)$ is the Pochhammer symbol, is ill-defined, as well as the gamma function in the second term. Instead, we regulate the expression by setting 
\begin{align}
    c = a + b + 1 + \epsilon
\end{align}
for an infinitesimally small $\epsilon$. Using 
\begin{align}
    \Gamma(-n-\epsilon) &= \frac{(-1)^n}{n!}\frac{1}{\epsilon}, \\
    \omega^\epsilon &= 1 + \epsilon\log \omega.\\
    \Gamma(\alpha + \epsilon) &= \Gamma(\alpha)(1 + \epsilon\psi(\alpha)),
\end{align}
where the limit as $\epsilon\rightarrow 0$ is implied, we expand~(\ref{eq:gauss_type_transform}) for $a=-\alpha^+, b=-\alpha^-, c=1-\alpha^+-\alpha^-$ and $\epsilon$. The simple poles in $\epsilon$ cancel between the two terms, and we are left with 
\begin{align}\label{eq:hulthen_small_r}
    G_0(r)+iF_0(r)&= C_0^{-1}e^{ipr}\bigg[\sum_{n=0}^\ell \frac{(-\alpha^+)_n(-\alpha^-)_n}{ n!(n-1)!}(1-e^{-m_\phi r})^{n} \nonumber \\
    &-\delta^{-1}\Big(\psi(1-\alpha^+) + \psi(1-\alpha^-) - \gamma_E + \log(1-e^{-m_\phi r})\Big)\nonumber \\
    &\phantom{=========}\times\sum_{m=0}^\infty \frac{(1-\alpha^+)_m(1-\alpha^-)_m}{(m+1)! m!}(1-e^{-m_\phi r})^{m+1}\bigg],
\end{align}
where $\gamma_E$ is the Euler-Mascheroni constant. 
From this, we can read off 
\begin{align}
    g_{0,1} &= -\frac{1}{\hat p C_0}\big(\Re\{\psi(1 - \alpha^+) + \psi(1 - \alpha^-)\} - \gamma_E - \log(p/m_\phi)\big) \\
    x_0(p) &= \left(\frac{p}{m_\phi}\right)^{-1}\frac{|\Gamma(1 - \alpha^+ - \alpha^-)|^2}{\Gamma(-\alpha^+)\Gamma(-\alpha^-)\Gamma(1+\alpha^+)\Gamma(1+\alpha^-)}
\end{align}
which we can use to evaluate the leading terms in $\bar Z_0$.

\subsubsection{Behavior near resonance}

Noting the form of~(\ref{eq:hulthen_sommfac}), we see that resonance occurs when the real parts of the arguments of the gamma functions are non-positive integers, since the Sommerfeld factor will go to infinity for vanishing $p$ (off-resonance, the enhancement is regulated by finite $m_\phi$ and does not diverge as $p\rightarrow 0$). Thus, resonance occurs when 
\begin{align}\label{eq:resonance_condition}
    \frac{2\mu\alpha}{m_\phi} = (n+1)^2.
\end{align}
For this value of $m_\phi$, and for small $p$, the Sommerfeld factor takes the form 
\begin{align}
C_0 &= \Bigg|\frac{\Gamma(-n - ip/m_\phi)\Gamma(n+2-ip/m_\phi)}{\Gamma(1 - 2i p/m_\phi)}\Bigg|\\
&= \sqrt {\frac{\pi}{2}\frac{m_\phi}{p}\frac{\sinh(2\pi p/m_\phi)}{\sinh^2(\pi p/m_\phi)}\prod_{k=1}^{n+1}(k^2 + (p/m_\phi)^2)^{-1}} \\
&\xrightarrow{p\to 0}
 (n+1)!\frac{m_\phi}{p},
\end{align}
which diverges as $p\rightarrow 0$.
Near resonances, the Sommerfeld factor will saturate at a value dependent on some ``distance'' in parameter space from resonance. We consider a small deviation $\varepsilon$ from the resonance condition~(\ref{eq:resonance_condition})
\begin{align}
    \frac{2\mu\alpha}{m_\phi} = (n+1)^2 + 2(n+1)\varepsilon,
\end{align}
where now $\varepsilon$ is a finite, dimensionless quantity,
and the factors multiplying $\varepsilon$ have been chosen for convenience. Expanding in $p,\varepsilon$, the Sommerfeld factor is
\begin{align}
C_0 &= \Bigg|\frac{\Gamma(-n - i\hat p (n+1)^2 - \varepsilon)\Gamma(n + 2  -i\hat p(n+1)^2+\varepsilon)}{\Gamma(1 - i\hat p (n+1)^2)} + \mathcal{O}(p^2, \varepsilon^2, p\varepsilon)\bigg|  \nonumber \\
&\simeq (n+1)!|\Gamma(-n - i\hat p (n+1)^2 - \varepsilon)(1 + (\varepsilon-i\hat p (n+1)^2 )\psi(n + 2)+i\hat p (n+1)^2 \gamma| \nonumber \\
&\simeq \frac{n+1}{\sqrt{\hat p^2 (n+1)^4 + \varepsilon^2}}.
\end{align}
where the last line holds for both small $p$ and $\varepsilon$. For $\varepsilon\rightarrow 0$ we recover the on-resonance result. For finite $\varepsilon$, the $ C_0$ saturates roughly when $\hat p\sim\varepsilon/(n+1)^2$. 

We can now consider the leading-order behavior of the PSS24 regulator. Plugging in our result for the power series of $G_0$ and expanding the digamma functions near resonance, we obtain
\begin{align}
\label{eq:PSS_regulator_hulthen}
    \bar Z_0&\simeq p\bigg(C_0g_{0,1} + \frac{2\mu\alpha}{p}\Big(1 + \log \frac{p}{\mu}\Big)\bigg) \\
    &=-2\mu\alpha\bigg(\Re\{\psi(1-\alpha^+) + \psi(1-\alpha^-)\} + \log \frac{e^{1-\gamma_E}m_\phi}{\mu_a}\bigg)\\
    &=-2\mu\alpha\bigg(\frac{\varepsilon}{\hat p^2 (n+1)^4 + \varepsilon^2} + \psi(n+2) + \psi(n+1) + \log \frac{e^{1-\gamma_E}m_\phi}{\mu_a}\bigg),
\end{align}
where we have used $\psi(-n -\varepsilon + i\hat p (n+1)^2)\approx \frac{1}{\varepsilon - i\hat p (n+1)^2} + \psi(n+1)$ in the third line. We see that precisely on-resonance, the outgoing-wave regulator is small, and negligible when multiplied by the $s$-wave annihilation amplitude $\bar f_{s,0}\sim \alpha^2/\mu$ (here $\alpha$ is the presumably weak coupling governing the annihilation physics). However, close to resonance, the regulator is dominated by the first term and saturates to a large value 
\begin{align}
    \bar Z_0&\xrightarrow{p\rightarrow 0}-\frac{2\mu\alpha}{\varepsilon},
\end{align}
which is not negligible in modifying the regulated cross sections.

Let us now consider the phase shift near a resonance for the purposes of evaluating the W25 regulator. Near resonance, we have 
\begin{align}
    \delta_0&=\arg\bigg\{\frac{\Gamma(1-\alpha^-)\Gamma(1-\alpha^+)}{\Gamma(1-\alpha^+-\alpha^-)}\bigg\}\\
    &\simeq  \arg\{-\varepsilon+i\hat p (n+1)^2 + \mathcal{O}(\varepsilon^2, \hat p\varepsilon, \hat p^2)\},
\end{align}
and so to leading-order in $p,\epsilon$ the W25 outgoing-wave regulator is 
\begin{align}
    pC_0^2\cot \delta_0 &\simeq p(n+1)^2\frac{1}{\hat p^2 (n+1)^2 + \varepsilon^2}\left(\frac{-\varepsilon}{\hat p (n+1)^2} \right)\nonumber \\&=-2\mu\alpha\frac{\varepsilon}{\hat p^2 (n+1)^4+\varepsilon^2}.
\end{align}
The first and last equalities above are understood to only hold for the large term that saturates at $1/\varepsilon$ for small $p$. The above is precisely the dominant term in the PSS24 regulator~(\ref{eq:PSS_regulator_hulthen}). The sub-leading terms in both regulators are small and negligible when combined with the perturbative annihilation amplitude $\bar f_{s,0}$. We therefore have that, in all of the on-resonance, near-resonance, and far-from-resonance cases, the unitarized $s$-wave cross sections obtained by the PSS24 and W25 regulating approaches considered here coincide for scattering via the long-range Hulth\'{e}n potential, as expected from our previous arguments.

\section{Regulating Sommerfeld resonances in multi-state systems}
\label{sec:multistate}
In this section, we generalize the results of Section~\ref{sec:sec_equiv_a_matching_RG_ss} to annihilation in systems with multiple coupled channels. We will find that, near a resonance, the large, $a$-independent term of the PSS24 regulator can be expressed solely in terms of quantities depending on the long-range physics. For simplicity, we assume that the states are coupled by a long-range, real and symmetric potential $V_{L,ij}(r)$, and have approximately the same mass $\mu$. Following Ref.~\cite{Parikh:2024mwa}, we define the asymptotic momenta of each of the $i=1,...,N$ states as $p_n=\sqrt{p^2-2\mu\lim_{r\rightarrow \infty}V_{L,nn}(r)}$. We take the first $M$ states to be kinematically open, and so $N-M$ of the $p_n$ are purely imaginary. In what follows, we assume that the masses of the mediators of the long-range forces are much smaller than $\mu$, so as previously there is a large hierarchy between $1/\mu$ and the range of the potential.

The annihilation cross section for a two-body system prepared in an initial state $i\in \{1,...,M\}$, where the short-distance annihilation physics is assumed to be entirely contained within $r<a$, is~\cite{Parikh:2024mwa}
\begin{align}
    (\sigma v_{\text{rel}})_{i, \text{ann}}=c_i\frac{2\pi i}{\mu}(2\ell+1)\Big[\Sigma_\ell^\dagger(\kappa_\ell^\dagger - \kappa_\ell)\Sigma_\ell\Big]_{ii} \label{eq:multistate_ann_xsec_old}
\end{align}
where 
\begin{align}
\label{eq:ms_definitions}
    \Sigma_\ell &= \Big[1-i\Sigma_{0,\ell}P\Sigma_{0,\ell}^\dagger\kappa_\ell\Big]^{-1}\Sigma_{0,\ell} \nonumber \\
    \kappa_\ell^{-1} &= \bar f_{s,\ell}^{-1}-\bar Z_\ell \nonumber \\
    \Sigma_{0,\ell}&= \tilde P Q_\ell P^{-1} \nonumber \\
    \Sigma_{0,\ell}^a&= \tilde P Q_\ell^a \tilde P^{-1} \nonumber \\
    \bar Z_\ell &= \Sigma_{0,\ell}^a\tilde P^{1/2}\alpha_{\tilde G_\ell}(a)(\tilde P^\dagger)^{-\ell}.
\end{align}
$\bar f_{s,\ell}$, now a matrix, is the purely short-distance scattering amplitude for the multi-state system, with the contributions already included in $V_L(r)$ factored out. $\Sigma_{0,\ell}$ is the standard multi-state Sommerfeld matrix, and $\Sigma_{0,\ell}^a$ is the Sommerfeld matrix for the same potential when the potential is set to zero outside the matching radius. Here, $\tilde P_{ij}=p_i\delta_{ij}$ is an $N\times N$ matrix of (possibly complex) momenta and $ P$ is its $M\times M$ real truncation (note this represents a difference in notation relative to Ref.~\cite{Parikh:2024mwa}). $Q_\ell$ is the $N\times M$ matrix defined as the boundary condition of the regular solution $w_\ell$ at the origin 
\begin{align}
    w_\ell\rightarrow  s_\ell(\tilde Pr)Q_\ell
\end{align}
whose $i$-th column corresponds to the regular solution for an incoming plane wave in the $i$-th channel. Here, by $s_\ell(\tilde{P} r)$ we mean the $N\times N$ diagonal matrix whose $(ij)$ entry is $s_\ell(p_i r)\delta_{ij}$; we will often denote this matrix simply by $s_\ell$ in what follows, and similarly for its analogue $c_\ell$.

$\tilde F_\ell$ and $\tilde G_\ell$ are $N\times N$ matrices whose columns describe  regular and irregular families of solutions, with each column defined to be the real ($\tilde{G}_\ell$) or imaginary ($\tilde{F}_\ell$) parts of a solution that is purely outgoing (or exponentially suppressed in the $N-M$ kinematically closed channels) at infinity, with short-distance boundary conditions
\begin{align}
    \tilde G_{\ell,ij}(r\rightarrow 0) &\rightarrow (2\ell-1)!!r^{-\ell}\delta_{ij} \label{eq:multistate_BCs_short} \\
    \tilde F_{\ell,ij}(r\rightarrow 0)&\rightarrow \frac{1}{(2\ell+1)!!}r^{\ell+1}z_{ij},
\end{align}
at the origin. Here, $z_{ij}$ describes a $N\times N$ matrix of finite values, and we can show that $z_\ell=\tilde P^{\ell+1}Q_\ell P^{-1}Q_\ell^\dagger (\tilde P^\dagger)^{\ell+1}$.  This result can be obtained by matching the short- and long-distance Wronskians between the regular and irregular families (see Ref.~\cite{Parikh:2024mwa} for details). 

Note that while $\tilde F_\ell$ is $N\times N$, it is only rank $M$, since there are only $M$ linearly independent regular solutions where the kinematically  forbidden channels are restricted to contain only exponentially decaying modes. In more detail, since there are $2N$ linearly independent solutions overall, and all regular solutions share $N$ common boundary conditions that impose regularity at the origin, the requirement of no kinematically decaying modes imposes another $N-M$ shared conditions, leaving $M$ degrees of freedom. In contrast, $\tilde G_\ell$ is rank $N$, since its short-distance boundary condition forces all $N$ of its columns to represent independent solutions.  

We define $\alpha$ and $\beta$ to be the $r$-dependent coefficients appearing in the variable phase decomposition 
\begin{align}
    u_\ell(r) = f_\ell(r)\alpha(r) - g_\ell(r)\beta(r),
\end{align}
where $f_\ell(r)=s_\ell(\tilde Pr)\tilde P^{-1/2}$ and $g_\ell(r)=(c_\ell(\tilde Pr) + is_\ell(\tilde Pr))\tilde P^{-1/2}$. As in the single-state case, we fix the variable phase solutions such that
\begin{align}
    u_\ell'(r) = f'_\ell(r)\alpha(r) - g'_\ell(r)\beta(r).
\end{align}

\subsection{Expressing the outgoing-wave regulator in terms of long-distance physics}
To express the regulated $S$-matrix of the PSS24 method in terms of the long-distance physics, we proceed in the same spirit as Section~\ref{sec:equivalence_singlestate}: we obtain an approximate expression for the scattering solutions near the range of the potential, and expand near the resonance to show that the regular solution is approximately related to the irregular solution by a constant matrix. Sufficiently near the resonance, this relation will hold also at the matching radius, allowing us to write the PSS24 regulator as a large, $a$-independent constant dependent only on the long-range physics. We will confirm this result with numerical examples. Appendix~\ref{app:multistateexpansion} contains expressions for the long-range scattering solutions in terms of similar solutions for a truncated short-range potential, and an alternative derivation for the constant matrix approximately relating the regular and irregular solutions (under the assumption, justified in the main text, that such a matrix exists).

First, we write the regulator $\bar Z_\ell$ in terms of the regular $\tilde F_\ell$ and irregular $\tilde G_\ell$ solutions and their Wronskians with the free-particle solutions. To find the Sommerfeld factor for when the potential is set to zero for  $r>a$, we define an additional basis of regular solutions $\bar F_\ell(r)$ with boundary conditions 
\begin{align}
    \bar F_\ell(r)\rightarrow s_\ell(\tilde Pr)
\end{align}
at the origin. Note that these solutions include the exponentially growing modes in kinematically forbidden channels and are therefore rank $N$.
The solutions outside the matching radius are that of a free (diagonal) unit-normalized plane wave plus a purely outgoing piece weighted by a scattering amplitude
\begin{align}
    u^a_{\ell,>}(r) = s_\ell(\tilde Pr) + (c_\ell(\tilde Pr)+ is_\ell(\tilde Pr))f_{\ell,0}^a\tilde P.
\end{align}
Taking the Wronskian of this wavefunction with the purely outgoing free-particle solution ($c_\ell + i s_\ell$) at $r=a$ picks out the coefficient of the $s_\ell$ plane-wave piece, which in matrix form is just the identity matrix, multiplied by a $\tilde{P}$ factor from differentiating the wavefunctions (which are free-particle solutions) with respect to $r$.

The solutions for $r\le a$ will be of the form $u_{\ell,<}^a(r)=\bar F_\ell (r)Q_\ell^a$. For our purposes, we need only the factor $Q_\ell^a$, which we can extract by:
\begin{align}
    Q_\ell^a=W_a[c_\ell + i s_\ell, \bar F_\ell]^{-1}\tilde P,
\end{align}
where $W_a[a_\ell,b_\ell] = a_\ell^T(a) b'_\ell(a) - a_\ell'^T(a) b_\ell(a)$ is the multi-state Wronskian matrix evaluated at $r=a$. 

It will be useful in the following to relate the quantity above to the regular solutions $\tilde F_\ell$. By comparing the boundary conditions at the origin, the solutions $\tilde F_\ell$ of the full scattering problem are related to $\bar F_\ell$ by 
\begin{align}
    \tilde F_\ell=\bar F_\ell Q_\ell P^{-1}Q_\ell^\dagger (\tilde P)^{\ell+1},
\end{align}
so that
\begin{align}
    W_a[c_\ell + i s_\ell, \bar F_\ell]^{-1}W_a[c_\ell+is_\ell,\tilde F_\ell]= Q_\ell P^{-1}Q_\ell^\dagger (\tilde P^\dagger)^{\ell+1}.
\end{align}

Using our expression for $Q_\ell^a$ and the definition of $\alpha_{\tilde G_\ell}$, the outgoing-wave regulator takes the form
\begin{align}
\label{eq:ZbarEll_ms_Wronskian}
    \bar Z_\ell = \tilde P W_a[c_\ell + is_\ell, \bar F_\ell]^{-1}W_a[c_\ell + is_\ell, \tilde G_\ell] (\tilde P ^\dagger)^{-\ell}. 
\end{align}
In what follows, we will show that near a resonance, the relevant parts of the solutions $\tilde F_\ell$ and $\tilde G_\ell$ within the range of the potential and sufficiently far from the origin are approximately related by a constant matrix $\tilde G_\ell\simeq\tilde F_\ell\gamma$. Note this cannot be an equality because the rank of the solutions $\tilde F_\ell$ is $M$ while $\tilde G_\ell$ is full-rank; however, it can be true that the corrections that encode the additional degrees of freedom are small in terms of their contribution to the outgoing-wave regulator between $r=a$ and $r=R$. When we can make this identification, the outgoing-wave regulator can be written 
\begin{align}
\label{eq:ZbarEll_ms_gamma}
    \bar Z_\ell = \tilde P Q_\ell P^{-1}Q_\ell^\dagger (\tilde P^\dagger)^{\ell+1}\gamma (\tilde P ^\dagger)^{-\ell},
\end{align}
which we will show is a  large, $a$-independent constant matrix.

To find the constant matrix relating the relevant parts of the regular and irregular solutions near a resonance, we consider the boundary conditions of the regular and irregular solutions at infinity, and continue the solutions into the range $R$ of the potential. Here we assume there is little evolution due to the long-range potential for $r>R$. Following the definitions of Sec. 3 in Ref.~\cite{Parikh:2024mwa}, as $r\rightarrow \infty$, the purely outgoing solution approaches 
\begin{align}
\label{eq:wtil_BCs_long}
    \tilde G_\ell(r) + i\tilde F_\ell(r) \rightarrow (c_\ell(\tilde{P} r) + i s_\ell(\tilde{P} r))DP^{-1}Q_\ell^T\tilde P^{\ell+1},
\end{align}
 where $D=(I_M \:\:0_{M\times(N-M)})^T$ is a $N\times M$ matrix that picks out the free-particle solutions in the $M$ kinematically allowed channels. Further, the regular $N\times M$ solution $w_\ell$ is related to $\tilde F_\ell$ by 
\begin{align}
\label{eq:tildeF_w_relation}
    \tilde F_\ell=w_\ell P^{-1}Q_\ell^\dagger (\tilde P^\dagger)^{\ell+1}
\end{align}
and has large-$r$ boundary conditions
\begin{align}
\label{eq:w_BCs_long}
    w_\ell(r\rightarrow \infty) \rightarrow s_\ell(\tilde{P} r) D + (c_\ell(\tilde{P} r) +is_\ell(\tilde{P} r))D f_{0,\ell}P,
\end{align}
where $f_{0,\ell}$ defines the long-range uncorrected $S$-matrix
\begin{align}
    S_{0,\ell}=1 + 2i\sqrt{P}f_{0,\ell}\sqrt{P}.
\end{align}
Using~(\ref{eq:tildeF_w_relation}) and the long-distance behaviors~(\ref{eq:w_BCs_long}),~(\ref{eq:wtil_BCs_long}), and equating the coefficients of $c_\ell$ and $s_\ell$ for the two solutions, the boundary conditions for $\tilde F_\ell$ and $w_\ell$ are related by
\begin{align}
    \text{Re}(DP^{-1}Q_\ell^T\tilde P^{\ell+1})&=D(1 + if_{0,\ell}P)P^{-1}Q_\ell^\dagger (\tilde P^\dagger)^{\ell+1} \\
    \text{Im}(DP^{-1}Q_\ell^T\tilde P^{\ell+1})&=Df_{0,\ell}Q_\ell^\dagger (\tilde P^\dagger)^{\ell+1}.
\end{align}
Note that the $D$ matrices ensure that only the kinematically allowed final states are relevant to this matching, and so the relevant entries of $c_\ell(\tilde{P} r)$ and $s_\ell(\tilde{P} r)$ are real.

Now, we use the definition of the $S$-matrix in terms of the scattering amplitude and the $M\times M$ $K$-matrix
\begin{align}
    S_{0,\ell} &=(1 + iK_\ell)(1 - iK_\ell)^{-1} \Leftrightarrow i K_\ell = (S_{0,\ell} - 1)(S_{0,\ell} + 1)^{-1}
\end{align}
to write the boundary conditions as  
\begin{align}
    \text{Re}(DP^{-1}Q_\ell^T\tilde P^{\ell+1})&=DP^{-1/2}\frac{1}{1-iK_\ell}P^{-1/2}Q_\ell^\dagger (\tilde P^\dagger)^{\ell+1} \\
    \text{Im}(DP^{-1}Q_\ell^T\tilde P^{\ell+1})&=DP^{-1/2}\frac{K_\ell}{1-iK_\ell}P^{-1/2}Q_\ell^\dagger (\tilde P^\dagger)^{\ell+1},
\end{align}
where the above will prove to be a convenient form for expanding near a resonance. Continuing the form~(\ref{eq:wtil_BCs_long}) into the range of the potential, the solutions therefore take the form
\begin{align}
\label{eq:ms_FG_at_R}
    \tilde F_\ell( R) &\simeq s_\ell(\tilde PR) DP^{-1/2}\frac{1}{1-iK_\ell}P^{-1/2}Q_\ell^\dagger (\tilde P^\dagger)^{\ell+1} + c_\ell(\tilde P R) DP^{-1/2}\frac{K_\ell}{1-iK_\ell}P^{-1/2}Q_\ell^\dagger (\tilde P^\dagger)^{\ell+1} \nonumber \\
    \tilde G_\ell( R) &\simeq c_\ell(\tilde P R) DP^{-1/2}\frac{1}{1-iK_\ell}P^{-1/2}Q_\ell^\dagger (\tilde P^\dagger)^{\ell+1} - s_\ell(\tilde P R) DP^{-1/2}\frac{K_\ell}{1-iK_\ell}P^{-1/2}Q_\ell^\dagger (\tilde P^\dagger)^{\ell+1}.
\end{align}

To understand the momentum scaling of~(\ref{eq:ms_FG_at_R}), we employ the multi-channel effective range expansion~\cite{Ross:1961jlg}
\begin{align}
\label{eq:ERE_ms}
P^{\ell+1/2}K_\ell^{-1}P^{\ell+1/2} = M_\ell + \frac{1}{2}R_\ell P^2 + \mathcal{O}(P^4),
\end{align} 
which is the multi-channel analog of~(\ref{eq:ERE_ss}). Here, $M_\ell$ is a matrix with small eigenvalues near a resonance, with at least one eigenvalue equal to zero on-resonance. For brevity, we write $E_\ell=M_\ell + \frac{1}{2}R_\ell P^2$, which is a matrix with small components near a resonance for small $P$. The solutions near the range of the potential can then be written 
\begin{align}
\label{eq:FG_at_R_ms}
    \tilde F_\ell(\tilde P R) &\simeq s_\ell(\tilde PR) DP^{-\ell-1} E_\ell (E_\ell-i P^{2\ell+1})^{-1}P^{\ell}Q_\ell^\dagger (\tilde P^\dagger)^{\ell+1} \nonumber\\
    &\phantom{================}+ c_\ell(\tilde P R) DP^{\ell}(E_\ell-i P^{2\ell+1})^{-1}P^{\ell}Q_\ell^\dagger (\tilde P^\dagger)^{\ell+1} \nonumber \\
    \tilde G_\ell(\tilde P R) &\simeq c_\ell(\tilde P R) DP^{-\ell-1}E_\ell (E_\ell-i P^{2\ell+1})^{-1} P^{\ell}Q_\ell^\dagger (\tilde P^\dagger)^{\ell+1} \nonumber\\
    &\phantom{================}- s_\ell(\tilde P R) DP^{\ell}(E_\ell-i P^{2\ell+1})^{-1}P^{\ell}Q_\ell^\dagger (\tilde P^\dagger)^{\ell+1}.
\end{align}
Modulo the universal factor of $Q_\ell^\dagger (\tilde P^\dagger)^{\ell+1}$, for small (but non-zero) $E_\ell, \tilde PR$, the first term in $\tilde F_\ell(R)$ scales like $P^\ell$, and the second term scales like $E_\ell^{-1}P^\ell$. Further, the first term in $\tilde G_\ell(R)$ scales like $P^{-\ell-1}$, while the second term scales like $P^{2\ell+1}E_\ell^{-1}P^\ell$. Thus, for small $E_\ell$ and sufficiently small $P$, both $\tilde F_\ell(R)$ and $\tilde G_\ell(R)$ are dominated by the $c_\ell(\tilde PR)$ terms, as in the single-state case. Thus, near-resonance for all $\ell$, and on-resonance for $\ell>0$, we can write
\begin{align}
\label{eq:tildeG_tildeF_relation}
    \tilde G_\ell(r) \simeq\tilde F_\ell(r) (\tilde P^\dagger)^{-\ell-1}Q_\ell(Q_\ell^\dagger Q_\ell)^{-1}P^{1/2} K_\ell^{-1} P^{-1/2}Q_\ell^\dagger (\tilde P^\dagger)^{\ell+1}
\end{align}
near the range of the potential. We comment below on the special case of the on-resonance behavior for $\ell=0$, since the $c_0$ does not necessarily dominate over the $s_0$ term in $\tilde G_0(R)$, and so the above relation does not hold (as in the single-state case).

Sufficiently near the resonance, when the deviations from~(\ref{eq:tildeG_tildeF_relation}) are small, the relation~(\ref{eq:tildeG_tildeF_relation}) approximately holds until $r=a$. One can see this also by performing an analogous expansion of $\bar Z_\ell$ as in~(\ref{eq:Z_ell_LO}), where it is straightfoward to show the coefficient of $r^{\ell+1}$ in the series expansion for $\tilde G_\ell$ provides the only (potentially) non-negligible contribution to (\ref{eq:ZbarEll_ms_Wronskian}) (which happens to be $a$-independent). As in section~\ref{sec:equivalence_singlestate}, the series expansions of the solutions can then be evaluated at $r=R$ and matched to the expansions~(\ref{eq:FG_at_R_ms}). This matching again demonstrates that, sufficiently close to resonance,~(\ref{eq:tildeG_tildeF_relation}) holds also at $r\sim a$. As in the single-state case, the matching described above also demonstrates that in the case of the $\ell=0$ resonance, if $M_0$ vanishes, the outgoing-wave regulator $\bar Z_0$ is small and negligible. We note we do not generally expect $M_0=0$ on-resonance, since resonance occurs when at least one, and usually one, of the eigenvalues of $M_0$ are zero. In this case, where $M_0$ is non-zero, the $c_0$ terms dominate for non-zero $P$, and the above relation holds. Since the steps described above are identical to the single-state case, only with series expansion coefficients replaced by matrices, we do not explicitly perform them here.

The dominant contribution to the PSS24 regulator is therefore a large, $a$-independent constant, depending only on the long-range scattering information. Substituting~(\ref{eq:tildeG_tildeF_relation}) into~(\ref{eq:ZbarEll_ms_gamma}), we obtain:
\begin{align}
\label{eq:Zbarelll_ms_approx}
    \bar Z_\ell \simeq \Sigma_{0,\ell} P^{1/2} K_\ell^{-1} P^{1/2}\Sigma_{0,\ell}^\dagger.
\end{align}
This immediately yields a simplified result for $\kappa_\ell$ in (\ref{eq:ms_definitions}) that can be substituted into the annihilation cross section in (\ref{eq:multistate_ann_xsec_old}). Furthermore, since the $S$-matrix can be written as \cite{Parikh:2024mwa}
\begin{align} S_\ell = S_{0,\ell} (1 + 2 i P^{1/2} \Sigma_{0,\ell}^\dagger \left[\kappa_\ell^{-1} - i \Sigma_{0,\ell} P \Sigma_{0,\ell}^\dagger\right]^{-1} \Sigma_{0,\ell} P^{1/2}), \end{align}
this modified $\kappa_\ell$ result can also be applied to the full regulated $S$-matrix. Elastic scattering cross sections from state $i$ to state $f$ can then be computed from the $S$-matrix as:
\begin{align}\sigma_{i\rightarrow f} = \frac{\pi}{p_i^2} \sum_\ell (2\ell+1) |(S_{\ell})_{fi} - \delta_{fi}|^2. \end{align}

In the following section we will apply our result for $\bar{Z}_\ell$ to the case of wino DM annihilation, and show numerically that it is in excellent agreement with the original PSS24 regulator near a resonance. 

\subsection{Example: wino dark matter}
We now compare the approximate expression for the multi-state outgoing-wave regulator~(\ref{eq:Zbarelll_ms_approx}) to the exact expressions obtained via the PSS24 method.
For the purposes of testing our results for multiple coupled channels, we study the case of wino DM annihilation. 

\subsubsection{Background}

The wino annihilation cross section was
analyzed in detail in Ref.~\cite{Parikh:2024mwa}, and provides a
benchmark for verifying that our W25-motivated approximation coincides with the exact PSS24 result when relevant for regulating the cross sections.

The pure wino consists of an $SU(2)_L$ triplet of Majorana fermions, which can be described as a Majorana fermion + a charged Dirac fermion,
\begin{equation}
    \chi = (\chi^0,\,\chi^{\pm})
\end{equation}
with a small radiative mass splitting $\Delta = m_{\chi^+} - m_{\chi^0}$
between the neutral and charged components. Here, $\chi^0$ is the DM candidate. In the two-particle sector,
the relevant states for the long-distance scattering problem are $\chi^0\chi^0$ and $\chi^+\chi^-$ in the notation of the Method-2 basis of Ref.~\cite{Beneke:2014gja}. The wino annihilation system therefore represents a coupled-channel annihilation problem with
$N=2$. For center-of-mass energies below the threshold,
\begin{equation}
    E < 2\Delta,
\end{equation}
only the neutral channel is kinematically open, which is the
relevant regime for wino annihilation in the Milky Way halo. In what follows, we consider the $N=2, M=1$ case for the standard wino annihilation, and artificially set $\Delta=0$ to study the case of $N=2, M=2$ coupled channel annihilation.

At leading order, the electroweak exchange induces the
long-range potential (e.g.~\cite{Hisano:2004ds})
\begin{equation}
    V(r) =
    \begin{pmatrix}
        0 & -\sqrt{2}\,\alpha_W\,\dfrac{e^{-m_W r}}{r} \\
        -\sqrt{2}\,\alpha_W\,\dfrac{e^{-m_W r}}{r} &
        \:\:\:2\Delta - \alpha\,\dfrac{1}{r}
        - \alpha_W c_W^2 \dfrac{e^{-m_Z r}}{r}
    \end{pmatrix},
\end{equation}
where $\alpha_W = g^2/(4\pi)$ and $c_W = \cos\theta_W$ are the usual Standard Model electroweak parameters.
At tree-level, there is no long-range potential in the neutral channel,
while the charged channel experiences both Coulomb and Yukawa interactions. For benchmarking purposes, we neglect the next-to-leading-order corrections to the potential (as described in \cite{Beneke:2019qaa,Urban:2021cdu}), and we take the long-range potential to be independent of the matching radius $a$.

For simplicity, we focus on $s$-wave and $p$-wave annihilation ($\ell=0,1$). The short-distance
annihilation amplitude defines a $2\times 2$ matrix $\bar{f}_{s,\ell}$ in
channel space, and we take (e.g.~\cite{Parikh:2024mwa})
\begin{align}
\bar f_{s,0}
&=
i \, \frac{\alpha_W^2}{2(2\mu)}
\begin{pmatrix}
1 & \frac{1}{\sqrt{2}} \\
\frac{1}{\sqrt{2}} & \frac{3}{2}
\frac{1}{\sqrt{2}} & \frac{3}{2} \end{pmatrix},
\\
\bar f_{s,1}
&= \frac{7}{54}\,\frac{\alpha_W^2}{(2\mu)^3}
\begin{pmatrix}
i\,|p_1|^2 & -\frac{1}{\sqrt{2}}\,|p_1 p_2| \\
\frac{1}{\sqrt{2}}\,|p_1 p_2| & \frac{3}{2}\, i\,|p_2|^2
\end{pmatrix}.
\end{align}

\subsubsection{Results for comparison of outgoing-wave regulators and cross sections}
\begin{figure}[t]
    \centering
    \includegraphics[width=\textwidth]{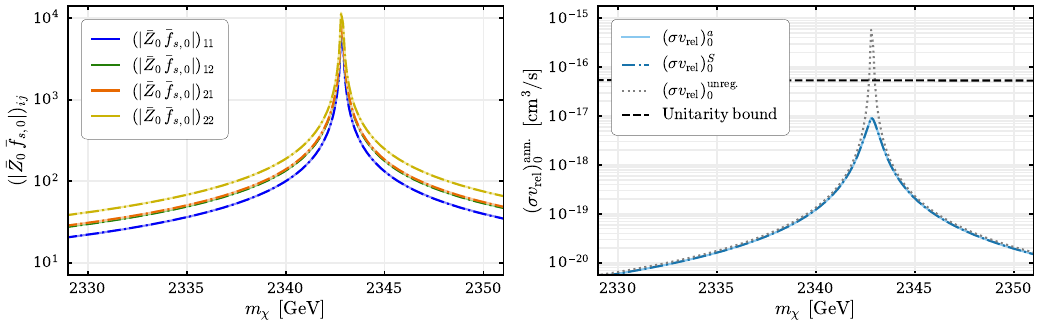}
    \caption{
    Absolute values of the matrix elements of 
    $\bar Z_0 \bar f_{s,0}$ (left) and the annihilation cross section (right) for wino annihilation at $v_{\rm{rel}}=10^{-6}$. The matrix elements of the exact PSS24 outgoing-wave regulator (lighter-shaded solid curves, left) coincide with those from the approximate result (darker-shaded dash-dot curves, left) to visual accuracy. The unregulated cross section (grey dotted, right) peaks above the unitarity bound sufficiently close to the resonance. The regulated cross sections obtained via the exact (light-teal solid, right) and approximate (dash-dotted teal, right) also coincide to visual accuracy.
    }
    \label{fig:wino_M=1_l=0_mchi_scan}
\end{figure}
In Figure~\ref{fig:wino_M=1_l=0_mchi_scan} we compare the absolute values of the matrix elements of
$\bar Z_0 \, \bar f_{s,0}$ computed using~(\ref{eq:Zbarelll_ms_approx})
to those obtained from the PSS24 expression~(\ref{eq:ms_definitions}). 
The matrix elements are plotted in the mass
interval $m_\chi \in [2330,2350]~\mathrm{GeV}$ for
$v_{\rm rel}=10^{-6}$, which is centered on the first $s$-wave resonance.
All four matrix elements exhibit the same characteristic resonant
profile, which is sharply peaked at the resonance with a steep symmetric falloff. The exact PSS24 result is plotted in lightly shaded solid colors and the approximate result is plotted in darker-shaded dash-dotted lines. In this mass window, the exact PSS24 and approximate forms for the outgoing-wave regulator coincide to within visual accuracy, and no relative shift of the resonance or distortion
of its shape is observed. The exact PSS24 cross section, denoted here $(\sigma v_{\rm rel})_{\ell=0}^a$, and the same cross section with the replacement~(\ref{eq:Zbarelll_ms_approx}), denoted here $(\sigma v_{\rm rel})_{\ell=0}^S$, therefore coincide to within visual accuracy and do not peak above the unitarity bound.

\begin{figure}[t]
    \centering
    \includegraphics[width=\textwidth]{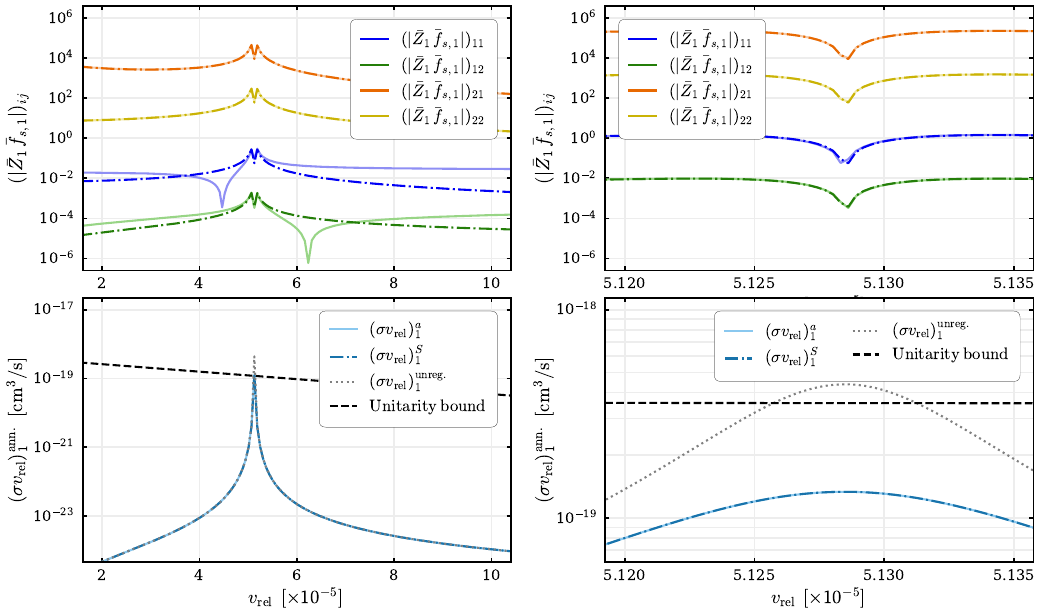}
    \caption{Absolute values of the matrix elements of 
    $\bar Z_1 \bar f_{s,1}$ (top panels) and the annihilation cross section (bottom panels) at the $m_\chi=11270.9999$ [GeV] $p$-wave super-resonance. The right panels are zoomed in relative to the left panels. The matrix elements of the exact PSS24 outgoing-wave regulator (lighter-shaded solid curves, top) coincide to visual accuracy with the matrix elements of the approximate expression (darker-shaded dash-dotted curves, top) when they are either large or very close to resonance. The exact PSS24 regulated cross section (solid teal, bottom) and the cross section regulated with the approximate regulator (dash-dotted darker teal, bottom) also coincide to visual accuracy, and do not peak above the unitarity bound. 
    }
    \label{fig:wino_M=1_l=1_vrel_scan}
\end{figure}

In Figure~\ref{fig:wino_M=1_l=1_vrel_scan} we plot the $p$-wave super-resonance of the wino system for $m_\chi=11270.9999$ GeV, which occurs at finite velocity $v_{\rm{rel}}\simeq5.1258\times 10^{-5}$. 
We observe a hierarchy between the elements of $\bar Z_1 \bar f_{s,1}$, i.e. some components are more relevant to the regulated cross section than others. The components $(\bar Z_1 \bar f_{s,1})_{ij}$ of the approximate result agree with those from the exact PSS24 calculation when the components are large compared to 1 (and therefore non-negligible in the cross section). When the components $(\bar Z_1 \bar f_{s,1})_{ij}$ are small, which is the case for the $(21)$ and $(22)$ components, the approximate result agrees with the exact result only very close to the resonance. Since this disagreement is negligible as a correction, the approximate and exact cross sections coincide to visual accuracy.  

\begin{figure}[t]
    \centering
    \includegraphics[width=\textwidth]{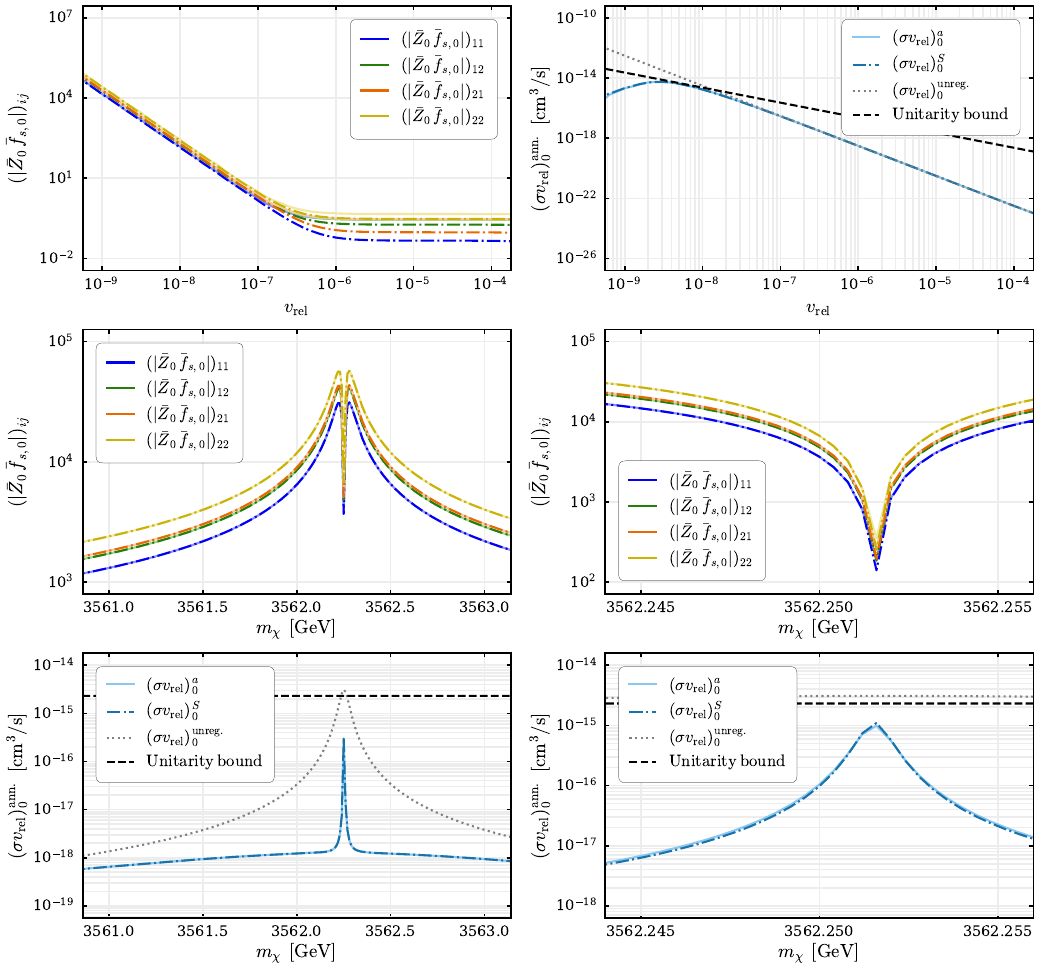}
    \caption{Absolute values of the matrix elements of 
    $\bar Z_0 \bar f_{s,0}$ (top left and middle panels) and the annihilation cross section (top right and bottom panels) for $s$-wave annihilation of $\chi^0\chi^0$ when the mass splitting $\Delta$ of the wino system is artificially set to zero. The top panels show the outgoing-wave regulator entries and annihilation cross section on the resonance for $m_\chi=3562.2516$ GeV. At $v_{\rm{rel}}\simeq 10^{-8}$, the regulated cross section (teal, top right) diverges from the unregulated cross section (grey dotted, top right), which is where the unregulated cross section exceeds the unitarity bound. The middle and bottom panels show the matrix elements of the outgoing-wave regulator and cross section, respectively, near the resonance for $v_{\rm{rel}}=10^{-8}$; the right panels are a zoom-in of the left panels. 
    }
    \label{fig:wino_M=2_l=0}
\end{figure}

We also see that, just as for the case of the Hulth\'{e}n potential, the regulator dips near the resonance due to the smallness of the eigenvalues of $M_\ell$ in~(\ref{eq:ERE_ms}). Note however, that for the multi-state case, not all the eigenvalues of $M_\ell$ are zero at a resonance, and so the approximate expression for the regulator does not vanish to $\mathcal O(P^4)$ near the resonance. 

In Figure~\ref{fig:wino_M=2_l=0}, we consider an $s$-wave resonance in the wino system, again for $\chi^0\chi^0$ annihilation, where we have artificially set $\Delta=0$ for the purposes of benchmarking our results for $M=2$ coupled channels near a resonance. The resonance occurs for $m_\chi=3562.2516$ GeV. In the top panels, we plot the elements $(\bar Z_0 \bar f_{s,0})_{ij}$ of the outgoing-wave regulator (top-left) and the annihilation cross section (top-right) in the velocity window $v_{\rm{rel}}\in[10^{-9}, 10^{-3}]$. We see that within this window, the approximate result for the unitarized cross section exactly matches the exact PSS24 result within visual accuracy. Further, the elements $(\bar Z_0 \bar f_{s,0})_{ij}$ match until approximately $v_{\rm{rel}}\simeq10^{-7}$, where they asymptote to quantities less than 1 for larger velocities. In the high velocity regime, the regulator is negligible and the cross sections coincide. In the low-velocity regime, the regulator scales like $p^{-2}$, as expected from~(\ref{eq:Z_ell_LO}). 

We also plot the elements $(\bar Z_0 \bar f_{s,0})_{ij}$ and the cross sections at fixed velocity $v_{\rm{rel}}=10^{-8}$ in the mass window $m_\chi\in[3561,3563]$ GeV. Near the resonance, the corrected cross section has a significantly narrower peak than the uncorrected cross section, and does not peak above the unitarity bound. We do not yet have a clear physical understanding of the remaining narrow peak in this $M=2$ case, except to note that it does not appear to be a numerical artifact (but may be an artifact of taking $\Delta=0$ but still specifying an initial state purely in one of the two degenerate two-particle states, which is unlikely to be a physically-realized initial condition).

\section{Conclusions}
\label{sec:conclusions}

We have examined three proposed methods for unitarizing the Sommerfeld enhancement in the single-channel, single-state case (where their assumptions overlap), and have found that they yield equivalent results. We have demonstrated that the full regulated $S$-matrix obtained using the results of PSS24 \cite{Parikh:2024mwa} can be matched exactly to that of FP25 \cite{Flores:2025uoh}, provided that (a) we assume the absorptive physics is confined to the region $r < a$, and (b) we match the short-range scattering amplitude $\bar{f}_{s,\ell}$ of PSS24 to a specific expression involving integrals of the wavefunctions and absorptive potential for $0 < r < a$ (given in (\ref{eq:FPmatching})). This expression makes manifest how the nominally $a$-dependent results of PSS24 emerge from the $a$-independent results of FP25; the short-range scattering amplitude is formally $a$-dependent (since it is defined in terms of scattering of a wave that is plane-wave-like at $r=a$), and this cancels the $a$-dependence in the regulator terms. The matching between $\bar{f}_{s,\ell}$ and the bare amplitudes of FP25 may involve UV-divergent terms, but we have demonstrated explicitly that in the PSS24 formalism, these terms are folded into the definition of $\bar{f}_{s,\ell}$ (as they are controlled by the UV physics, corresponding to the limit as $r\rightarrow 0$, which should be computed using a renormalized UV theory), and are separated from the IR enhancement due to the long-range potential. However, in cases beyond the scope of PSS24, where there is no hierarchy of scales between the absorptive physics and the long-range potential, there may be non-trivial interplay between these effects for which the formalism of FP25 is needed.

To perform the comparison to W25 \cite{Watanabe:2025kgw}, we have demonstrated that close to resonances, the irregular and purely outgoing wavefunction behaves very similarly to the regular wavefunction between the matching radius and the range of the potential. This is a consequence of the effective range expansion and is independent of the details of the potential. One of the terms in the equation for the regulated Sommerfeld enhancement in PSS24 involves the properties of the purely outgoing wavefunction, as the effect of the short-range physics is to source an additional outgoing (scattered) wave; this relationship allows us to rewrite this ``outgoing-wave regulator'' solely in terms of the regular solution, resulting in an expression that is manifestly independent of the matching radius and agrees with W25. This approximation is only valid close to resonances, but this is also where the correction factor is large, and we have checked in both an analytically-tractable example and a realistic case that this approximate solution agrees very closely with the exact result from PSS24. 

Furthermore, we have demonstrated that this argument can be extended to the multi-state case studied in PSS24, leading to a new simplified expression for the regulated Sommerfeld enhancement. Essentially, if $\bar{f}_{s,\ell}$ is the scattering amplitude describing the short-range physics in the absence of the long-range potential (including annihilation and any short-range scattering not captured in the potential), the corrected Sommerfeld factor to be contracted with the annihilation matrix can be written as:
\begin{align} \Sigma_\ell = \left[ 1 - \Sigma_{0,\ell} P^{1/2} (i +  K_\ell^{-1}) P^{1/2} \Sigma_{0,\ell}^\dagger \bar{f}_{s,\ell} \right]^{-1} \Sigma_{0,\ell},
\end{align}
where $\Sigma_{0,\ell}$ is the usual unregulated Sommerfeld enhancement matrix, $P$ is a diagonal matrix whose entries are the state momenta $p_i$, and $K_\ell$ is the scattering $K$-matrix involving only the long-range potential. The full $S$-matrix, which can be used to compute the regulated results for elastic scattering as well as annihilation, is given by:
\begin{align} S_\ell = S_{0,\ell} \left(1 + 2 i P^{1/2} \Sigma_{0,\ell}^\dagger \left[\bar{f}_{s,\ell}^{-1} -\Sigma_{0,\ell} P^{1/2} ( i + K_\ell^{-1}) P^{1/2} \Sigma_{0,\ell}^\dagger\right]^{-1} \Sigma_{0,\ell} P^{1/2}\right), \end{align}
where $S_{0,\ell}$ is the $S$-matrix in the absence of the short-range physics described by $\bar{f}_{s,\ell}$.

\section*{Acknowledgments} 

The authors thank Kalliopi Petraki, Marcos Flores,  Tobias Binder, and Yuki Watanabe for helpful discussions. This work was supported by the U.S. Department of Energy, Office of Science, Office of High Energy Physics of U.S. Department of Energy under grant contract Number DE-SC0012567. 
BC is supported by the Natural Sciences and Engineering Research Council of Canada through a Canada Graduate Scholarship --- Doctoral (Grant No. 599773).
TRS' work is supported by the Simons Foundation (Grant Number 929255, T.R.S); during the course of this work, T.R.S.~was also supported in part by a Guggenheim Fellowship; the Edward, Frances, and Shirley B.~Daniels Fellowship of the Harvard Radcliffe Institute; and the Bershadsky Distinguished Fellowship of the Harvard Physics Department. 

\begin{appendix}
\section{Series expansions of $F_\ell$ and $G_\ell$} \label{sec:detail on Fell and Gell}
\label{app:analyticwavefn} 

The behavior of the radial wavefunctions near the origin is determined by the singular structure of the radial Schr\"odinger equation
\begin{align}
\left[-\partial_r^2+\frac{\ell(\ell+1)}{r^2}+2\mu V_L(r)-p^2\right]u_\ell(r)=0.
\end{align}
The centrifugal term (and possibly the potential) render the equation singular at $r=0$. The local behavior of the solutions may be obtained using the Frobenius method near the origin. Assuming the potential is no more singular than $r^{-1}$ at the origin, the regular and irregular solutions have the forms 
\begin{align}
    F_\ell(r)\sim r^{\ell+1}, \qquad G_\ell\sim r^{-\ell},
\end{align}
near the origin in general. 

For small $r$, we expand the solutions in power series of $x=pr$,
\begin{align}
\label{eq:power_seriesFG}
F_\ell(r) &= \sum_{k=\ell+1}^{\infty} f_{\ell,k} x^k , \\
G_\ell(r) &= \sum_{k=-\ell}^{\infty} g_{\ell,k} x^k + x_\ell(p)\log xF_\ell(r) .
\end{align}
Here, the logarithmic term in the irregular solution is required for the Wronskian between the two solutions to be constant. More generally, such logarithmic contributions may appear whenever the indicial roots of the singular differential equation differ by an integer. In the limit of vanishing potential, these functions reduce to the spherical Bessel functions of the first and second kind, where we choose the normalizations
\begin{align}\label{eq:power_seriessc}
F_\ell &\to pr\, j_\ell(pr)
       =\frac{x^{\ell+1}}{(2\ell+1)!!}
        +\mathcal O(x^{\ell+3}), \\
G_\ell &\to -pr\, y_\ell(pr)
       =(2\ell-1)!!\,x^{-\ell}
        +\mathcal O(x^{-\ell+2}),
\end{align}
where the arrow denotes the limit of vanishing  long-range potential.

We apply the Schr\"odinger equation to the power series above to obtain the recursion relations between their coefficients. Writing the potential as
\begin{align}
V_L(r)=\sum_{m=-1}^\infty V_m r^m ,
\end{align}
and substituting into
\begin{align}
\left[-\partial_x^2+\frac{\ell(\ell+1)}{x^2}+\frac{2\mu}{p^2} V(r)-1\right]u_\ell=0 ,
\end{align}
one finds for the regular solution 
\begin{align}
\bigl[\ell(\ell+1)-k(k-1)\bigr] f_{\ell,k}
+ \frac{2\mu}{p^2}\sum_{m=-1}^{k-\ell-2} V_m\, p^{-m}\,
f_{\ell,k-m-2}
- f_{\ell,k-2}
=0 ,
\qquad k\ge \ell+1 .
\label{eq:Frobenius_regular_recursion_k}
\end{align}
Equivalently, for \(k\neq \ell+1\),
\begin{align}
\label{eq:recursion_flk}
f_{\ell,k}
=
\frac{
\displaystyle
\frac{2\mu}{p^2}\sum_{m=-1}^{k-\ell-2} V_m\, p^{-m}\,
f_{\ell,k-m-2}
- f_{\ell,k-2}
}{
k(k-1)-\ell(\ell+1)
}.
\end{align}

For the irregular solution, the substitution yields
\begin{align}
\label{eq:recursion_glk}
&\bigl[\ell(\ell+1)-k(k-1)\bigr] g_{\ell,k}
+ \frac{2\mu}{p^2}\sum_{m=-1}^{k+\ell-2} V_m\, p^{-m}\,
g_{\ell,k-m-2}
- g_{\ell,k-2}
 \nonumber \\
 &\phantom{=======================}+ x_\ell(p)\,(2k-1)\,f_{\ell,k}
=0 ,
\qquad k\ge -\ell ,
\end{align}
where the last term contributes only when \(k\ge \ell+1\), since \(f_{\ell,k}=0\) for \(k<\ell+1\).

For \(k=\ell+1\), the prefactor of \(g_{\ell,\ell+1}\) vanishes, and the consistency condition determines the logarithmic coefficient \(x_\ell(p)\):
\begin{align}
\frac{2\mu}{p^2}\sum_{m=-1}^{2\ell-1} V_m\, p^{-m}\,
g_{\ell,\ell-m-1}
- g_{\ell,\ell-1}
+ (2\ell+1)x_\ell(p)\,f_{\ell,\ell+1}
=0 .
\label{eq:Frobenius_log_constraint_k}
\end{align}
Equivalently,
\begin{align}
\label{eq:recursion_xl}
x_\ell(p)
=
\frac{
\displaystyle
g_{\ell,\ell-1}
-\frac{2\mu}{p^2}\sum_{m=-1}^{2\ell-1} V_m\, p^{-m}\,
g_{\ell,\ell-m-1}
}{
(2\ell+1)f_{\ell,\ell+1}
}.
\end{align}
Note the coefficients $f_{\ell,\ell+1}=C_\ell/(2\ell+1)!!$ and $g_{\ell,-\ell}=C^{-1}_\ell(2\ell-1)!!$ are fixed by normalizations~(\ref{eq:power_seriessc}). 

\section{Power series expansion of $\bar Z_\ell$}
\label{app:exp_barZell}
To understand the short-distance behavior of $\bar Z_\ell$ it will be useful for the following to expand the Wronskians between $\{F_\ell,G_\ell\}$ and $\{s_\ell,c_\ell\}$ in terms of their series expansions at the origin. Using~(\ref{eq:power_seriesFG}) and the series expansions
\begin{align}
s_\ell(x) &= x^{\ell+1}\sum_{k=0}^{\infty}
\frac{(-\tfrac12 x^2)^k}{k!(2\ell+2k+1)!!}, \\
c_\ell(x) &= x^{-\ell}\left[
\sum_{k=0}^{\infty}\frac{(2\ell-2k-1)!!}{k!}
\left(\tfrac12 x^2\right)^k
+(-1)^\ell\sum_{k=\ell+1}^{\infty}
\frac{(-\tfrac12 x^2)^k}{k!(2k-2\ell-1)!!}
\right],
\end{align}
we compute the Wronskians $\bar W_x[f,g]=f(x)g'(x)-f'(x)g(x)=p^{-1}W_r[f,g]$ as power series in $x=pr$. Expanding to leading order yields the relations given below:
\begin{align}
W_x[s_\ell, F_\ell] 
&= \sum_{n=0}^\infty \frac{(-\tfrac{1}{2} x^2)^n}{n! (2k + 2n + 1)!!}\sum_{k=\ell+1}^\infty k \, f_{\ell, k} \, x^{k-1} \nonumber \\
&\phantom{====}-  \sum_{n=0}^\infty \frac{(2n + \ell + 1)}{n! (2k + 2n + 1)!!} \left( -\tfrac{1}{2} x^2 \right)^n \sum_{k=\ell+1}^\infty f_{\ell, k} \, x^k \, x^2\nonumber \\
&= \sum_{m=\ell+1}^\infty \sum_{n=0}^\infty 
\left( \frac{m - (\ell+1) - 4n}{n! (2\ell + 2n + 1)!!} \right)
(-\tfrac{1}{2})^n f_{\ell, m - 2n} \, x^{2\ell + m} 
\nonumber \\
&= \frac{1}{(2\ell + 1)!!} f_{\ell, \ell+2} \, x^{2\ell + 2} + \mathcal{O}(x^{2\ell + 3})
\end{align}

\begin{align}
W_x[c_\ell, F_\ell] 
&= 
\left[
\sum_{n=0}^\infty \frac{(2\ell - 2n - 1)!!}{n!} \left( \tfrac{1}{2} x^2 \right)^n 
+ (-1)^\ell \sum_{n=\ell+1}^\infty \frac{(-\tfrac{1}{2} x^2)^n}{n! (2n - 2\ell - 1)!!}
\right]\sum_{k=\ell+1}^\infty k f_{\ell,k} \, x^{k-1} \, x^{-\ell} \nonumber \\
&\phantom{\!\!\!\!\!\!\!\!\!\!\!\!\!\!\!\!\!\!\!\!\!}
- 
\Bigg[
\sum_{n=0}^\infty \frac{(2n - \ell)(2\ell - 2n - 1)!!}{n!} \left( \tfrac{1}{2} x^2 \right)^n + (-1)^\ell \sum_{n=\ell+1}^\infty \frac{(2n - \ell)(-\tfrac{1}{2} x^2)^n}{n! (2n - 2\ell - 1)!!}
\Bigg]\sum_{k=\ell+1}^\infty f_{\ell,k} \, x^k \, x^2 x^{-\ell-1} \nonumber \\
&= \sum_{m=\ell+1}^\infty \sum_{n=0}^\infty
\frac{(m - 4n + \ell)(2\ell - 2n - 1)!!}{n!} 
\left( \tfrac{1}{2} \right)^n f_{\ell, m - 2n} \, x^{m - (\ell+1)} \nonumber \\
&\phantom{===} + (-1)^\ell \sum_{m=2(\ell+1)}^\infty \sum_{n=\ell+1}^\infty 
\frac{(m - 2n + \ell)(2\ell - 2n - 1)!!}{n!} 
\left( \tfrac{1}{2} \right)^n f_{\ell, m - 2n} \, x^{m - (\ell+1)} \nonumber \\
&= C_\ell + (2\ell + 2)(2\ell - 1)!! \, f_{\ell,\ell+2} \, x + \mathcal{O}(x^2)
\end{align}

\begin{align}
W_x[s_\ell, G_\ell]&= \sum_{n=0}^\infty \frac{(-\tfrac{1}{2} x^2)^n}{n! (2k + 2n + 1)!!} \Bigg(
\sum_{k=\ell}^\infty k \, g_{\ell,k} \, x^{k-1}
+ x_\ell(p) \sum_{k=\ell+1}^\infty f_{\ell,k} \, x^{k-1}
\nonumber \\
&\phantom{==========================}+ x_\ell(p) \log x \sum_{k=\ell+1}^\infty k f_{\ell,k} \, x^{k-1}
\Bigg) \nonumber \\
&\phantom{+} - x^2 \sum_{n=0}^\infty \frac{(2n + \ell + 1)}{n! (2k + 2n + 1)!!} (-\tfrac{1}{2} x^2)^n\left(
\sum_{k=\ell}^\infty g_{\ell,k} x^k
+ x_\ell(p) \log x \sum_{k=\ell+1}^\infty f_{\ell,k} x^k
\right)  \nonumber \\
&= \sum_{m=\ell+1, n=0}^\infty \frac{(m - 4n + \ell)}{n! (2\ell + 2n + 1)!!}
(-\tfrac{1}{2})^n g_{\ell, m - 2n} \, x^{m - 2n}
\nonumber \\
&\qquad + x_\ell \log x \sum_{m=\ell+1, n=0}^\infty \frac{(m - (\ell+1) - 4n)}{n! (2\ell + 2n + 1)!!}
(-\tfrac{1}{2})^n f_{\ell, m - 2n} \, x^{m + \ell} \nonumber \\
&\qquad + x_\ell\sum_{m=\ell+1, n=0}^\infty \frac{(\frac{-1}{2})^n}{n!(2\ell+1+2n)!!}f_{\ell, m-2n}x^{m+\ell} \nonumber \\
&= C_\ell^{-1} + \frac{1}{(2\ell - 1)!!} \, g_{\ell,\ell+1} \, x^{\ell-1} + \mathcal{O}(x^2)
\nonumber \\
&\qquad + x_\ell \log x \left( \frac{1}{(2\ell + 1)!!} f_{\ell,\ell+2} \, x^{2\ell + 2}
+ \mathcal{O}(x^{2\ell+3}) \right) \nonumber \\
&\qquad +x_\ell\left( \frac{C_\ell}{(2\ell+1)!!}x^{2\ell+1} + \mathcal{O}(x^{2\ell+1})\right)
\end{align}

\begin{align}
W_x[G_\ell, c_\ell] 
&= \left(
\sum_{k=\ell}^\infty k \, g_{\ell,k} \, x^{k-1}
+ x_\ell(p) \sum_{k=\ell+1}^\infty f_{\ell,k} \, x^{k-1}
+ x_\ell(p) \log x \sum_{k=\ell+1}^\infty k f_{\ell,k} \, x^{k-1}
\right) x^{-\ell} \notag \\
&\qquad \times \left[
\sum_{n=0}^\infty \frac{(2\ell - 2n - 1)!!}{n!} \left( \tfrac{1}{2} x^2 \right)^n
+ (-1)^\ell \sum_{n=\ell+1}^\infty \frac{(-\tfrac{1}{2} x^2)^n}{n! (2n - 2\ell - 1)!!}
\right] \notag \\
&\qquad - \left(
\sum_{k=\ell}^\infty g_{\ell,k} \, x^k
+ x_\ell(p) \log x \sum_{k=\ell+1}^\infty f_{\ell,k} \, x^k
\right) x^{2} x^{-\ell -1} \notag \\
&\qquad \times \left[
\sum_{n=0}^\infty \frac{(2n - \ell)(2\ell - 2n - 1)!!}{n!} \left( \tfrac{1}{2} x^2 \right)^n
+ (-1)^\ell \sum_{n=\ell+1}^\infty \frac{(2n - \ell)(-\tfrac{1}{2} x^2)^n}{n! (2n - 2\ell - 1)!!}
\right] \notag \\
&= \sum_{m=\ell, n=0}^\infty \frac{(m + n - \ell)(2\ell - 2n - 1)!!}{n!} \left( \tfrac{1}{2} \right)^n g_{\ell, m + 2n} \, x^{m - \ell} \notag \\
&\qquad + (-1)^\ell \sum_{m=2\ell+1, n=\ell+1}^\infty \frac{(m + n - \ell)(2\ell - 2n - 1)!!}{n!} \left( \tfrac{1}{2} \right)^n g_{\ell, m + 2n} \, x^{m - \ell} \notag \\
&\qquad + x_\ell(p) \log x \sum_{m=\ell+1, n=0}^\infty \frac{(m - n + \ell)(2\ell - 2n - 1)!!}{n!} \left( \tfrac{1}{2} \right)^n f_{\ell, m - 2n} \, x^{m - (\ell+1)} \notag \\
&\qquad + x_\ell(p) \bigg(
\sum_{m=\ell+1, n=0}^\infty \frac{(2\ell - 2n - 1)!!}{n!} \left( \tfrac{1}{2} \right)^n f_{\ell, m - 2n} \, x^{m - (\ell+1)}
\notag \\
&\quad\quad\quad\quad + (-1)^\ell \sum_{m=2\ell+1, n=\ell+1}^\infty \frac{(2\ell - 2n - 1)!!}{n!} \left( \tfrac{1}{2} \right)^n f_{\ell, m - 2n} \, x^{m - (\ell+1)}\bigg) \notag \\
&= 
\begin{cases}
g_{0,1} + x_0 C_0 + x_0 C_0 \log x + \mathcal{O}(x), & \ell = 0 \\
(2\ell - 1)!! g_{\ell, -\ell + 1} \, \big(x^{-2\ell} + \mathcal{O}(x^{-\ell + 2})\big) + (2\ell+1)!!g_{\ell,\ell+1}\big(1 + \mathcal{O}(x^{2}) \big) \\
\phantom{=====}+ x_\ell C_\ell( (2\ell+1)^{-1}+\log x), & \ell>0
\end{cases}
\end{align}

To justify the series approximation to the Wronskians for small enough $x$, it is sufficient to examine the recursion relations for the coefficients~(\ref{eq:recursion_flk}),~(\ref{eq:recursion_glk}), and~(\ref{eq:recursion_xl}), noting the powers of $x$ they are accompanied by. For a potential with $V_{-1}=-\alpha$, for example, it is straightforward to obtain the leading terms 
\begin{align}
x_0 C_0^2 &= -\frac{2\mu\alpha}{ p}, \\
g_{\ell,-\ell+1} &= \frac{(2\ell-1)!!}{\ell}\frac{\mu\alpha}{ p C_\ell}.
\end{align}
More generally, the coefficients $f_{\ell,k}$ and $g_{\ell,k}$ contain powers of $p$ that are equal to or higher than those appearing in $f_{\ell,\ell+1}$ and $g_{\ell,-\ell}$. In the expression for $\bar Z_\ell$, these higher-order coefficients always appear multiplied by additional powers of $pa$. As a result, they contribute only subleading corrections in the small-$pa$ expansion. An important exception is $g_{\ell,\ell+1}$, which is fixed by the boundary condition of $G_\ell$ at infinity, and can be understood as controlling the amplitude of the regular part of the irregular solution. We do not assume this term is suppressed relative to the leading-order coefficients, and it will turn out to give the dominant contribution to $\bar Z_\ell$ near a resonance. For the purpose of determining the leading behavior of $\bar Z_\ell$, we therefore retain only the coefficients $f_{\ell,\ell+1}$, $g_{\ell,-\ell}, x_\ell$, and $g_{\ell,\ell+1}$. 

Given the Wronskian relations above, we now determine the leading behavior of $\bar Z_\ell$ in the single-state case. We substitute the expansions for the Wronskians above into
\begin{align}\label{eq:Wronskian_Zbarell}
\bar Z_\ell
= p C_\ell^2
\frac{W_a[c_\ell+i s_\ell, G_\ell]}
{W_a[c_\ell+i s_\ell, F_\ell]} .
\end{align}
to obtain
\begin{align}\label{eq:barZell_LO}
\bar Z_\ell =
\begin{cases}
pC_0 g_{0,1}-2\mu\alpha\log x+\ldots , & \ell=0 ,\\[6pt]
\dfrac{[(2\ell-1)!!]^2}{\ell}\alpha\mu x^{-2\ell}
+(2\ell+1)!!\,pC_\ell g_{\ell,\ell+1}+px_\ell C_\ell^2\log x +\ldots , & \ell>0 ,
\end{cases}
\end{align}
where $x=pa$ and the ellipses denote the subleading contributions in the expansion.

We saw in the main text that the $g_{\ell,\ell+1}$ term dominates the expansion above. In special cases it is possible that $g_{\ell,\ell+1}=0$, such as the case when there is no potential. Here, it is the $g_{\ell,\ell+2}$ term that controls the weight of the regular solution in the irregular solution and will be the dominant term in the above expansion. For the classes of potentials we consider, however, the coefficient $g_{\ell,\ell+1}$ is generally non-zero. 

It will also be useful to note the scaling of the $g_{\ell,\ell+1}$ and $x_\ell C_\ell^2$ terms. Noting the form~(\ref{eq:power_seriesFG}), we must have that $C_\ell g_{\ell,\ell+1}$ scales at least as strongly as $C_\ell g_{\ell,\ell+1}\sim p^{-2\ell-1}$ with $p$ for small $p$. To see this, consider $p^\ell C_\ell G_\ell(r)$, which scales as $r^{-\ell}$ for small $r$. $p^\ell C_\ell G_\ell(r)$ is also a solution to the Schr\"odinger equation, and so will generally have a non-zero coefficient of $r^{\ell+1}$ in its series expansion. The coefficient $p^{2\ell+1}C_\ell g_{\ell,\ell+1}$ of $r^{\ell+1}$ will therefore generally take a finite value as $p\rightarrow 0$, and so $C_\ell g_{\ell,\ell+1}\sim p^{-2\ell-1}$ for small $p$. Further, from the solution for $x_\ell$,~(\ref{eq:recursion_xl}), we see that $x_\ell$ is a weighted sum of the coefficients $g_{\ell,k\le\ell}$, and the leading-$p$ scaling of $x_\ell$ is $p^{-2\ell-1}C_\ell^{-2}$, so $x_\ell C_\ell^2\sim p^{-2\ell-1}$. The $g_{\ell,\ell+1}$ and $x_\ell C_\ell^2$ terms therefore appear comparable in the expansion~(\ref{eq:barZell_LO}). However, unlike $g_{\ell,\ell+1}$, $x_\ell$ is insensitive to the resonant behavior~(\ref{eq:ERE_ss}) since it is fixed by~(\ref{eq:recursion_xl}). As seen when matching the dominant terms in the expansion for $G_\ell(R)$ in~(\ref{eq:FG_at_V_range}), we must have $C_\ell g_{\ell,\ell+1}\sim p^{-4\ell-2}\tan\delta_\ell$, and so $C_\ell g_{\ell,\ell+1}$ will be enhanced by a factor $1/\epsilon$ or $1/p^2$ relative to $x_\ell C_\ell^2$ near- or on-resonance, respectively.

\section{Expansion of irregular and regular solutions in multi-state case for $r < a$}
\label{app:multistateexpansion}

In the single-state case, we found it convenient at various points to expand the wavefunctions $G_\ell(r)$ and $F_\ell(r)$ in terms of the analogous wavefunctions for the problem where $V_L(r)$ is set to zero outside $r=a$, in order to disentangle long-range and short-range physics. In this appendix we present the corresponding relations for the multi-state case.

It is convenient in this case to work with the $N\times M$ regular solution $w_\ell(r)$ and the $N\times N$ irregular solution $\tilde{w}_\ell(r) = \tilde{G}_\ell(r) + i \tilde{F}_\ell(r)$. The analogous solutions where $V_L(r)=0$ for $r > a$ can be labeled $w_\ell^a(r)$ and $\tilde{w}_\ell^a(r)$. However, we will be interested in basis solutions that have exponentially growing modes in kinematically forbidden states at $r=a$, since solutions to the full problem may have non-zero overlap with such modes at $r=a$ but still evolve to physically valid solutions at large $r$ due to the effects of $V_L(r)$. Thus we will take $w_\ell^a(r)$ to be $N\times N$ rather than $N\times M$, with additional columns corresponding to initial plane waves in kinematically forbidden states. We can then write:
\begin{align} \tilde{w}_\ell & =  \tilde{w}_\ell^a \cdot A_1 +  w_\ell^a \cdot B_1, \quad w_\ell  =  \tilde{w}_\ell^a \cdot A_2 +  w_\ell^a \cdot B_2\end{align}
where $A_1$ and $B_1$ are $N\times N$, and $A_2$ and $B_2$ are $N\times M$. 
As in the main text we will employ the $N\times M$ projection matrix $D$ to project onto the kinematically allowed two-particle states.

The boundary conditions of the irregular solutions are fixed at the origin (\ref{eq:multistate_BCs_short}), implying that $A_1=1$, $A_2=0$. Furthermore, the matching to free-particle propagation at $r=a$ implies the following conditions on the variable-phase coefficients of $\tilde{w}_\ell^a$, $w_\ell^a$ at $r=a$:
\begin{align} \alpha_{\tilde{w}_\ell^a}(a) = 0, \quad \alpha_{w_\ell^a}(a) = \tilde{P}^{1/2}. \end{align}
Thus we have $B_1 = \tilde{P}^{-1/2} \alpha_{\tilde{w}_\ell}(a)$, $B_2 = \tilde{P}^{-1/2} \alpha_{w_\ell}(a)$, and overall we obtain, for $r < a$:
\begin{align} \tilde{w}_\ell(r) & =  \tilde{w}_\ell^a(r) +  w_\ell^a(r) \tilde{P}^{-1/2} \alpha_{\tilde{w}_\ell}(a), \quad w_\ell(r)   =   w_\ell^a(r) \tilde{P}^{-1/2} \alpha_{w_\ell}(a). \end{align}

Now Ref.~\cite{Parikh:2024mwa} derived a result that in our notation takes the form:
\begin{align}\alpha_{w_\ell}(a) = \tilde{P}^{-1/2} (\Sigma_{0,\ell}^a)^{-1} \tilde{P} Q_\ell = \tilde{P}^{-1/2} (\Sigma_{0,\ell}^a)^{-1} \Sigma_{0,\ell} P.  \end{align} 
From that work we also have $\alpha_{\tilde{G}_\ell}(a) = \alpha_{\tilde{w}_\ell}(a) - i \alpha_{w_\ell}(a) \Sigma_{0,\ell}^\dagger (\tilde{P}^\dagger)^{\ell}$, and since we know $\bar{Z}_\ell = \Sigma^a_{0,\ell} \tilde{P}^{1/2} \alpha_{\tilde{G}_\ell}(a) (\tilde{P}^\dagger)^{-\ell}$, it follows that:
\begin{align} \alpha_{\tilde{w}_\ell}(a) & = i \tilde{P}^{-1/2} (\Sigma_{0,\ell}^a)^{-1} \Sigma_{0,\ell} P \Sigma_{0,\ell}^\dagger (\tilde{P}^\dagger)^\ell + \tilde{P}^{-1/2} (\Sigma^a_{0,\ell})^{-1} \bar{Z}_\ell (\tilde{P}^\dagger)^{\ell} \nonumber \\
& = \tilde{P}^{-1/2} (\Sigma_{0,\ell}^a)^{-1} \left[ i   \Sigma_{0,\ell} P \Sigma_{0,\ell}^\dagger  +   \bar{Z}_\ell \right] (\tilde{P}^\dagger)^{\ell} \end{align}

Thus our decomposition of $w_\ell(r)$ and $\tilde{w}_\ell(r)$ for $r < a$ can be written as:
\begin{align} \tilde{w}_\ell(r) & =  \tilde{w}_\ell^a(r) +  w_\ell^a(r) \tilde{P}^{-1} (\Sigma_{0,\ell}^a)^{-1} \left[ i   \Sigma_{0,\ell} P \Sigma_{0,\ell}^\dagger  +   \bar{Z}_\ell \right] (\tilde{P}^\dagger)^{\ell} , \nonumber \\
w_\ell(r) &  =   w_\ell^a(r) \tilde{P}^{-1}  (\Sigma_{0,\ell}^a)^{-1} \Sigma_{0,\ell} P. \end{align}

Up to this point, our results have been exact, but we would now like to understand how this expansion simplifies in the near-resonance region, in analogy to the single-state case. In the multi-state case, the $M$ columns of $w_\ell(r)$ span the space of regular solutions with the property that the components of the wavefunction corresponding to kinematically forbidden states decay exponentially as $r\rightarrow \infty$. We can divide $\tilde{w}_\ell(r)$ into a (regular) term determined by a linear combination of solutions stored in $w_\ell(r)$ (which can be expressed in the form of $w_\ell(r) C$ for some $M \times N$ matrix $C$), plus the irregular term (matching onto $\tilde{w}_\ell^a(r)$ at $r=a$) which ensures the correct short-distance BCs, plus an additional regular term which cancels the exponentially growing modes from the irregular term (and so cannot be written as $w_\ell(r) C$, which contains no exponentially growing modes). We may regard this last term as being a linear combination of the regular solutions with a unit-normalized ``incoming plane wave'' component in the kinematically forbidden states (which corresponds to an exponentially growing mode for imaginary momentum); the normalization of these terms is thus fixed completely by the size of the exponentially growing modes from the irregular term that need to be canceled.

If $\bar{Z}_\ell + i \Sigma_{0,\ell} P \Sigma_{0,\ell}^\dagger$ is sufficiently large, then the irregular term sourced at $r=a$ by $\tilde{w}_\ell^a(r)$ may be subdominant for $a \lesssim r \lesssim R$, and consequently the additional regular term that cancels out its exponentially growing components may also be neglected in this region. (In the main text, we give a more in-depth argument from the effective range expansion that this limit should hold sufficiently close to resonances and at low momentum.) If this limit indeed holds, then it implies that there exists a $M\times N$ matrix $C$ such that $w_\ell(r) C \approx \tilde{w}_\ell(r)$, satisfying:
\begin{align} \Sigma_{0,\ell} P C \approx \left[ i   \Sigma_{0,\ell} P \Sigma_{0,\ell}^\dagger  +   \bar{Z}_\ell \right] (\tilde{P}^\dagger)^{\ell} \Rightarrow \bar{Z}_\ell = \Sigma_{0,\ell}P \left[-i \Sigma_{0,\ell}^\dagger + C (\tilde{P}^\dagger)^{-\ell}\right] \end{align}
If this relation holds out to the range of the potential, where the variable-phase coefficients stop evolving, then we can also write:
\begin{align} \beta_{\tilde{w}_\ell}(r\rightarrow \infty)\approx \beta_{w_\ell}(r\rightarrow \infty) C.\end{align}
Using relations from Ref.~\cite{Parikh:2024mwa}, we can write $\beta_{w_\ell}(r\rightarrow \infty)$ in terms of the elastic scattering amplitude from the long-range potential, $D^T \beta_{w_\ell}(r\rightarrow \infty) = P^{1/2} f_{0,\ell} P$, and the projection onto the kinematically allowed states of $\beta_{\tilde{w}_\ell}(r\rightarrow \infty)$ satisfies:
\begin{align} D^T \beta_{\tilde{w}_\ell}(r\rightarrow \infty) =   P^{1/2} \Sigma_{0,\ell}^T  \tilde{P}^{\ell}.\end{align} 
Thus projecting onto the kinematically allowed states, we obtain:
\begin{align} P^{1/2} \Sigma_{0,\ell}^T \tilde{P}^\ell \approx P^{1/2} f_{0,\ell} P C \Rightarrow C \approx P^{-1} f_{0,\ell}^{-1} \Sigma_{0,\ell}^T \tilde{P}^\ell \end{align}
Using the relationship $S_{0,\ell} = 1 + 2 i P^{1/2} f_{0,\ell}P^{1/2}$, and also the identity $ \Sigma_{0,\ell}^T (\tilde{P}^{-1} \tilde{P}^\dagger)^{-\ell} = P^{-1/2} S_{0,\ell} P^{1/2} \Sigma_{0,\ell}^\dagger$ \cite{Parikh:2024mwa}, we can then write:
\begin{align}\bar{Z}_\ell & \approx \Sigma_{0,\ell}P \left[-i \Sigma_{0,\ell}^\dagger + P^{-1} f_{0,\ell}^{-1} \Sigma_{0,\ell}^T (\tilde{P}^{-1} \tilde{P}^\dagger)^{-\ell} \right] \nonumber \\
& = \Sigma_{0,\ell}P \left[-i  + 2 i P^{-1/2} (S_{0,\ell} - 1)^{-1} S_{0,\ell} P^{1/2}  \right] \Sigma_{0,\ell}^\dagger \nonumber \\
& =i  \Sigma_{0,\ell}P^{1/2} (S_{0,\ell} - 1)^{-1} \left[-(S_{0,\ell} - 1)  + 2   S_{0,\ell}   \right] P^{1/2} \Sigma_{0,\ell}^\dagger \nonumber \\
& = i  \Sigma_{0,\ell}P^{1/2} (S_{0,\ell} - 1)^{-1} (S_{0,\ell} +1) P^{1/2} \Sigma_{0,\ell}^\dagger \nonumber \\
& =  i  \Sigma_{0,\ell}P^{1/2} K_\ell^{-1} P^{1/2} \Sigma_{0,\ell}^\dagger \end{align}

\section{Extension of the FP25-PSS24 matching to the multiple-channel case}
\label{app:multichannel}

In the main text we focused on the case where there is only a single annihilation channel, and so the functions $v_\ell(r)$ governing the absorptive potential in the FP25 formalism are scalars. However, the full FP25 formalism allows those functions to carry a channel index $i$, such that $W_\ell$ becomes a matrix, and works out the exclusive cross sections corresponding to annihilation into different channels. In this appendix we work out how that formalism translates into the PSS24 approach. Note this ``multi-channel'' case is distinct from the ``multi-state'' case we consider in Section~\ref{sec:multistate}; in the latter case, there are multiple coupled initial two-body states with different masses, such that their initial momenta are distinct. In the multi-channel case of FP25, there is only one two-particle initial state, but multiple operators/channels through which the particles in that initial state may scatter or annihilate. 

The general expression for $W_\ell$ in FP25 (translated into our notation), employing the expressions for $F_\ell(r)$ and $G_\ell(r)$ from (\ref{eq:matching}), is given by:
\begin{align} W^{ij}_\ell(p) & =\frac{1}{p} \int^\infty_0 r dr \int^\infty_0 r^\prime dr^\prime F_{\ell}(r_<) G_{\ell}(r_>)v^i_\ell(r) v_\ell^{j*}(r') \nonumber \\
& \approx\frac{1}{p} \int^a_0 r dr \int^a_0 r^\prime dr^\prime (C_\ell/C^a_\ell) F^a_\ell(r_<) \left(\frac{C^a_\ell}{C_\ell} (G^a_\ell(r_>) + i F^a_\ell(r_>)) \right. \nonumber \\
& \left. + \frac{C_\ell}{C_\ell^a} \frac{\alpha_{G_\ell}(a)}{\alpha_{F_\ell}(a)} F^a_\ell(r_>) \right) v^i_\ell(r) v_\ell^{j*}(r') \end{align}
As previously we have assumed here that due to the $v^i_\ell(r)$ factors, we can truncate the integral outside $r=a$ (i.e.~all annihilation channels correspond to short-range physics). Generalizing our previous results, we can define the short-range integrals:
\begin{align} \Lambda^{ij}(a) & = \frac{\eta_\ell^i \eta_\ell^j}{p^2 (C^a_\ell)^2} \int^a_0 r dr \int^a_0 r' dr' F^a_\ell(r) F^a_\ell(r') v^i_\ell(r) v_\ell^{j*}(r'), \nonumber \\
\Lambda_1^{ij}(a) & = \frac{\eta_\ell^i \eta_\ell^j}{p} \int^a_0 r dr \int^a_0 r^\prime dr^\prime F^a_\ell(r_<) (G^a_\ell(r_>) + i F^a_\ell(r_>)) v^i_\ell(r) v_\ell^{j*}(r') \end{align}
Then we can write:
\begin{align} W^{ij}_\ell(p) 
& = \frac{1}{\eta_\ell^i \eta_\ell^j} \left[  \Lambda_1^{ij}(a) + p C_\ell^2 \frac{\alpha_{G_\ell}(a)}{\alpha_{F_\ell}(a)} \Lambda^{ij}(a) \right] = \frac{1}{\eta_\ell^i \eta_\ell^j} \left[  \Lambda_1^{ij}(a) + \bar{Z}_\ell \Lambda^{ij}(a) \right] \end{align}
Accordingly, $(\eta_\ell W_\ell \eta_\ell)^{ij} = \eta_\ell^i W_\ell^{ij} \eta_\ell^j = \Lambda_1^{ij}(a) + \bar{Z}_\ell \Lambda^{ij}(a)$, and in matrix form we can write $\eta_\ell W_\ell \eta_\ell = \Lambda_1(a) + \bar{Z}_\ell \Lambda(a)$, where $\bar{Z}_\ell$ is a scalar (defined as previously) and the $\Lambda$ functions are matrices encoding the short-range physics.

It then follows from FP25 that:
\begin{align} \eta_\ell M^\text{reg.}_\ell & = \left[1 -  \Lambda_1(a) - i \eta_\ell M_{\ell}^{\text{unreg.}} M_{\ell}^{\text{unreg.} \dagger} \eta_\ell  - \bar{Z}_\ell \Lambda(a) \right]^{-1} \eta_\ell M^\text{unreg.}_\ell \end{align}

Now using results from FP25 we can write the unregulated matrix elements as:
\begin{align} M^{\text{unreg.}, i}_{\ell} =\frac{1}{\sqrt{p}} i^\ell e^{i\delta_\ell}\int^a_0 F_\ell(r) v^i_\ell(r) = \frac{1}{\sqrt{p}} i^\ell e^{i\delta_\ell} (C_\ell/C^a_\ell) \int^a_0 F^a_\ell(r) v^i_\ell(r), \end{align}
and consequently:
\begin{align} (\eta_\ell M_{\ell}^{\text{unreg.}} M_{\ell}^{\text{unreg.} \dagger} \eta_\ell)^{ij} & = \eta^i_\ell M_{\ell}^{\text{unreg.} i} M_{\ell}^{\text{unreg.} j *}  \eta_\ell^j =  p C_\ell^2 \Lambda^{ij}(a), \end{align} 
so we can write:
\begin{align}\eta_\ell M^\text{reg.}_\ell & = \left[1 -  \Lambda_1(a) - (i p C_\ell^2  + \bar{Z}_\ell) \Lambda(a) \right]^{-1} \eta_\ell M^\text{unreg.}_\ell \end{align}
In the case where we work only at leading order in the pure-UV physics, we can drop the $\Lambda_1(a)$ term as higher-order in the short-range potential (compared to 1); if it is divergent, we take this step in the understanding that $\Lambda_1(a)$ represents a purely UV divergence that will be canceled when the UV theory is appropriately renormalized (akin to tree-level matching when the one-loop terms contain UV divergences). Then at leading order, where we can associate $\bar{f}_{s,\ell} \approx \Lambda(a)$, we expect the overall corrected cross section to be obtained by contracting the vector of uncorrected amplitudes with the prefactor matrix  $\left[1 -  (i p C_\ell^2  + \bar{Z}_\ell) \Lambda(a) \right]^{-1} $; this matches the prescription in PSS24 \cite{Parikh:2024mwa} for the exclusive cross sections (which is also computed only at leading order in the UV physics).

\end{appendix}

\bibliography{references}
\bibliographystyle{JHEP}

\end{document}